\PassOptionsToPackage{table,dvipsnames}{xcolor}
\documentclass[manuscript,screen]{acmart}
\AtBeginDocument{%
  \providecommand\BibTeX{{%
    Bib\TeX}}}

\setcopyright{acmlicensed}
\setcopyright{rightsretained}
\copyrightyear{2026}
\acmYear{2026}
\acmDOI{XXXXXXX.XXXXXXX}
\acmISBN{978-1-4503-XXXX-X/2018/06}

\usepackage{algorithmicx}
\usepackage{algorithm}
\usepackage{algpseudocode}
\algrenewcommand\alglinenumber[1]{\tiny #1}

\usepackage{fontawesome5}
\usepackage{forest}
\usepackage{tikz}
\usepackage{pgfplots}

\usepackage[textsize=tiny]{todonotes}
\def\done#1{}
\def\wontfix#1{}

\usepackage{subcaption}
\usepackage{graphicx}
\usepackage{textcomp}
\usepackage{listings}
\usepackage{hyperref}
\usepackage[capitalise, nameinlink, noabbrev]{cleveref}
\usepackage{relsize} 
\usepackage{xspace}
\usepackage[inline]{enumitem}

\usepackage[normalem]{ulem}
\usepackage{amsthm}
\usepackage{wrapfig}
\usepackage{alltt}

\usepackage{circledsteps}
\let\circled=\Circled

\newcommand{\conclusion}[1]{%
  \begin{center}
    \fbox{\begin{minipage}{0.75\columnwidth}
      \centering
      \emph{#1}
    \end{minipage}}
  \end{center}
}

\usepackage{tikz}
\usetikzlibrary{automata, positioning, arrows.meta, arrows}
\usepackage[frozencache,cachedir=.]{minted2} 
\usepackage{syntax} 
\newenvironment{densegrammar}{%
\begin{footnotesize}
\grammarparsep 3pt
\parsep 0pt
\itemsep 0pt
\begin{grammar}%\raggedright
}%
{%
\end{grammar}
\end{footnotesize}
}
\definecolor{DarkRed}{rgb}{0.6,0.0,0.0}
\definecolor{DarkBlue}{rgb}{0.0,0.0,0.6}
\def\server{\color{DarkRed}}
\def\client{\color{DarkBlue}}
\def\any{\color{black}}

\def\nonterm#1{\textnormal{$\langle$\emph{#1}$\rangle$}}

\newtheorem{definition}{Definition}

\usepackage{pgf-umlsd}

\def\BibTeX{{\rm B\kern-.05em{\sc i\kern-.025em b}\kern-.08em
    T\kern-.1667em\lower.7ex\hbox{E}\kern-.125emX}}

\def\smaller{\relscale{0.9}}

\def\AFL{{\smaller AFL}\xspace}
\def\AFLNET{{\smaller AFLNET}\xspace}

\def\ANTLR{{\smaller ANTLR}\xspace}
\def\API{{API}\xspace} 

\def\ChatGPT{{\smaller ChatGPT}\xspace}
\def\DNS{{\smaller DNS}\xspace}

\def\FANDANGO{{\smaller FANDANGO}\xspace}
\def\FTP{{\smaller FTP}\xspace}
\def\FuzztructionNET{Fuzztruction-{\smaller NET}\xspace}
\def\GRAMMARINATOR{{\smaller GRAMMARINATOR}\xspace}
\def\HTTP{{\smaller HTTP}\xspace}

\def\RTAG{{\smaller RTAG}\xspace}
\def\IO{interaction\xspace} 
\def\IIO{Interaction\xspace} 
\def\IP{{\smaller IP}\xspace}

\def\JSON{{\smaller JSON}\xspace}

\def\LLMs{{\smaller LLM}s\xspace}
\def\NuSMV{{\textsc{NuSMV}}\xspace}
\def\OPENAI{{OpenAI}\xspace}

\def\PEACH{{\smaller PEACH}\xspace}

\def\RFC{{\smaller RFC}\xspace}
\def\REST{{\smaller REST}\xspace}
\def\RESTAPI{{\smaller REST~API}\xspace}
\def\SMTP{{\smaller SMTP}\xspace}
\def\SPIN{{\smaller SPIN}\xspace}
\def\StateAFL{State{\smaller AFL}\xspace}
\def\JSON{{\smaller JSON}\xspace}
\def\TAMARIN{{\smaller TAMARIN}\xspace}
\def\TTCN{{\smaller TTCN}\xspace}
\def\UNIX{{\smaller UNIX}\xspace}

\def\ASCII{{\smaller ASCII}\xspace}
\def\XDR{{\smaller XDR}\xspace}
\def\XML{{\smaller XML}\xspace}
\def\FTPS{{\smaller FTPS}\xspace}
\def\TLS{{\smaller TLS}\xspace}

\def\IG{\text{IG}}

\newcommand{\crlf}{\char92r\char92n}

\begin{document}

\title{Language-Based Protocol Testing}

\author{Alexander Liggesmeyer}
\orcid{0009-0001-9778-3798}
\email{alexander.liggesmeyer@cispa.de}
\affiliation{%
 \institution{CISPA Helmholtz Center for Information Security}
 \city{Saarbrücken}
 \country{Germany}}

\author{José Antonio Zamudio Amaya}
\orcid{0000-0002-5025-7424}
\affiliation{%
 \institution{CISPA Helmholtz Center for Information Security}
 \city{Saarbrücken}
 \country{Germany}}
\email{jose.zamudio@cispa.de}

\author{Andreas Zeller}
\orcid{0000-0003-4719-8803}
\affiliation{%
 \institution{CISPA Helmholtz Center for Information Security}
 \city{Saarbrücken}
 \country{Germany}}
\email{andreas.zeller@cispa.de}

\renewcommand{\shortauthors}{Liggesmeyer, Zamudio Amaya, and Zeller}

\begin{abstract}
Over the past decade, the automated generation of test inputs has made significant advances.
Modern fuzzers and test generators easily produce complex input formats that do systematically cover the input and execution space.
Testing \emph{protocols,} though, has remained a frontier for automated testing, as a test generator has to \emph{interact} with the program under test, producing messages that conform to the current state of the system.
Common test generators that \emph{mutate} recorded interactions often fail at this point, as messages to be generated syntactically and semantically depend on messages received earlier.
Alternatively, one could model the protocol interactions.
However, existing \emph{state models} typically abstract away concrete protocol details to facilitate symbolic reasoning---details yet needed for testing concrete implementations.

In this paper, we introduce \emph{language-based protocol testing}, the first approach to specify, automatically test, and systematically cover the full state and input space of protocol implementations.
We specify protocols as \emph{\IO grammars}---an extension of context-free grammars that tag each message element with the communication party that is in charge of producing it.
\IIO grammars embed classical state models by unifying states, messages, and transitions all into nonterminals, and can be used for \emph{producing} interactions as well as \emph{parsing} them, making them ideally suited for testing protocols.
Additional \emph{constraints} over grammar elements allow us to specify and test \emph{semantic features} such as binary message formats, checksums, encodings, and the many ways that message features induce states and vice versa.

To evaluate the effectiveness of language-based protocol testing, we have implemented it as part of the \FANDANGO test generator.
We specify several protocols as \IO grammars, including features such as human-readable interactions (\SMTP), bit-level encodings (\DNS), and dynamic port assignments (\FTP), and use them to test the corresponding protocol implementations.
By systematically covering the \IO grammar and solving the associated constraints, \FANDANGO achieves comprehensive coverage of the protocol interactions, resulting in high code coverage and a thorough assessment of the program under test.
\end{abstract}



\received{26 February 2026}

\maketitle

\section{Introduction}
\label{sec:intro}

In the past decade, \emph{automated test generation} has become one of the fastest-growing fields of software engineering~\cite{potuzak2023current}.
This is mostly due to the success of \emph{fuzzing tools,}~\cite{weissberg2024sok} which produce massive amounts of random inputs often guided by coverage of the code and/or the input space.
Today, fuzzing tools are widely used to assess the robustness of systems~\cite{johansson2014t}.
They have become the prime tool to detect vulnerabilities in software:
Google's OSS-Fuzz project, running fuzzers 24/7, has helped identify and fix over 13,000 vulnerabilities and 50,000 bugs across 1,000 open source projects~\cite{google2025ossfuzz}.

\iffalse
Any kind of input with complex semantic constraints poses huge challenges for fuzzing.
Fuzzers typically operate by applying generic \emph{mutations} to existing inputs, such as flipping bits or exchanging bytes, and for complex inputs, this typically means that the input becomes \emph{invalid.}
Such limitations can be mitigated by providing a \emph{specification} of the input format~\cite{godefroid2008grammar}.
Context-free \emph{grammars,} for instance, are well-established formalisms to specify the syntax of strings; modern \emph{language-based} fuzzers combine such grammars with \emph{constraints}~\cite{zamudio2025fandango, dominicACM} to also cover \emph{semantic} features.
\fi

\def\exchange#1#2{\begin{call}{Client}{\client \textbf{#1}}{Server}{\server \textbf{#2}}\end{call}}
\begin{wrapfigure}[17]{R}{0.45\linewidth}
  
    \centering
    \resizebox{\linewidth}{!}{

    \begin{sequencediagram}\ttfamily\small
    \newthread{Client}{\textnormal{\small \client client}}{}
    \newinst[6cm]{Server}{\textnormal{\small \server server}}
    \exchange{(connect)}{220 smtp.example.com ESMTP Postfix}
    \exchange{HELO relay.example.org}{250 Hello relay.example.org, glad to meet you}
    \exchange{MAIL FROM:<bob@example.org>}{250 Ok}
    \exchange{RCPT TO:<alice@example.com>}{250 Ok}
    \exchange{DATA}{354 End data with <CR><LF>.<CR><LF>}
    \postlevel
    \postlevel
    \exchange{\shortstack{
        From: "Bob Example" <bob@example.org> \\
        To: "Alice Example" <alice@example.com> \\
        Subject: Protocol Testing with \IIO Grammars \\
        (mail body) \\
        .
        }
        }{250 Ok}
  \end{sequencediagram}
 }
\vspace{-2\baselineskip}
\caption{A simple \SMTP interaction}
\label{fig:smtp-sequence}
\end{wrapfigure}

\iffalse
However, ensuring the validity of individual inputs is only half the battle.
\fi
Despite these successes, there are still important frontiers for fuzzing.
One such frontier is \emph{protocol testing}~\cite{bochmann1994protocol}. 
Unlike stateless processing, where inputs are independent, protocol fuzzing requires the test generator to \emph{interact} with the program under test over time. 
The validity of a message depends not just on its syntax, but on the current \emph{state} of the communication.

As an example, consider the \SMTP~\cite{rfc-smtp} protocol, which is used to send email messages.
\SMTP is stateful; the client and server exchange messages in a specific order.
As shown in \cref{fig:smtp-sequence}, the client first connects to the server, then sends a \texttt{HELO} message, followed by a \texttt{MAIL FROM} message, and so on.
The \emph{protocol} (in this case, \SMTP) defines which messages are allowed in which state.
Since existing (recorded) interactions typically only cover a fraction of the possible interactions, and since mutating interactions often makes them invalid,
comprehensive coverage of such protocols almost universally requires a \emph{protocol specification.}

A classic way to specify a protocol is a \emph{state model}~\cite{cleeremans1989finite}---a finite state automaton (also called a \emph{labeled transition system}) in which state transitions are labeled with messages exchanged.
There is a vast field of methods available for specifying and checking stateful systems, and for symbolically \emph{reasoning} about protocols, all built around state models.
\Cref{fig:smtp-lts} shows such a state model for \SMTP. 

However, if we want to \emph{test} protocol implementations, an abstract specification is not sufficient. 
Our specification must be precise enough to be instantiated into \emph{concrete} messages and interactions.
We could, for instance, specify the format of messages as \emph{regular expressions}~\cite{polo2020automated}.
Then, {\server \nonterm{server:id}} could be defined as \verb!"220 .*\r\n"!--specifying a line starting with \texttt{220}, followed by arbitrary characters, and finally {\smaller CR} and {\smaller NL} line delimiters; such a regular expression can be easily instantiated into concrete matching strings.

\begin{figure*}
    \resizebox{\linewidth}{!}{
    \tikzset{
        ->,
        >=triangle 60,
        node distance=4cm,
        every state/.style={thick, fill=gray!10, minimum width=2cm},
        every edge/.style={draw, line width=2pt, gray!50, ->},
        initial text=$ $,
        }
    \begin{tikzpicture}\LARGE
        \node[state, initial] (start) {\nonterm{start}};
        \node[state, right of=start] (connect) {\nonterm{connect}};
        \node[state, right of=connect] (helo) {\nonterm{helo}};
        \node[state, right of=helo] (from) {\nonterm{from}};
        \node[state, right of=from] (to) {\nonterm{to}};
        \node[state, right of=to] (data) {\nonterm{data}};
        \node[state, accepting, right of=data] (quit) {\nonterm{quit}};
        \draw (start) edge[bend left, above] node{\color{black} (connect)} (connect)
              (connect) edge[bend left, above] node{\server \nonterm{server:id}} (helo)
              (helo) edge[bend left, above] node{\shortstack{\client \nonterm{client:HELO} \\ \server \nonterm{server:hello}}} (from)
              (from) edge[bend left, above] node{\shortstack{\client \nonterm{client:MAIL_FROM} \\ \server \nonterm{server:ok}}} (to)
              (to) edge[bend left, above] node{\shortstack{\client \nonterm{client:RCPT_TO} \\ \server \nonterm{server:ok}}} (data)
              (to) edge[loop above] node{\shortstack{\client \nonterm{client:RCPT_TO} \\ \server \nonterm{server:ok}}} (to)
              (data) edge[bend left, above] node{\shortstack{\client \nonterm{client:DATA} \\ \server \nonterm{server:ok}}} (quit)
              (helo) edge[bend right, below] node{\server \nonterm{server:error}} (quit)
              (from) edge[bend right, below] (quit)
              (to) edge[bend right, below] (quit)
              (data) edge[bend right, below] (quit)
        ;
    \end{tikzpicture}
    }
    \vspace{-1.5\baselineskip}
    \caption{A state model for the \SMTP protocol, specifying possible sequences of {\client client} and {\server server} messages. The concrete messages exchanged have been replaced with symbolic placeholders such as {\server \nonterm{server:id}} or {\client \nonterm{client:HELO}} to indicate which party is sending the message. The model does not specify the actual content or format of these messages.}
    \label{fig:smtp-lts}
\end{figure*}

\iffalse
Such a regular expression could even be translated into \emph{another} finite state automaton, further specifying the messages sent or expected; randomly traversing this automaton would allow us to produce various instantiations of the regular expression.
\fi
For specifying real-world message formats, however, regular languages are often too limited.
\emph{E-mail addresses}, for instance, can contain nested matching parentheses; their syntax thus cannot be expressed in a regular expression alone.
The same applies if we want to exchange \JSON messages (context-free) or \XML data (context-free for a given schema).
Such data formats must be specified as a \emph{grammar,}
consisting of \emph{rules} that expand \emph{nonterminals} into alternative sequences of terminal and nonterminal symbols.
Grammars are the oldest formalism for specifying languages~\cite{panini500bce}; as \emph{producers,} they form the oldest technique for generating test inputs for software~\cite{burkhardt1967,purdom1972}
Today, a wide array of techniques is available for instantiating grammars into concrete conforming strings~\cite{hodovan2018grammarinator, godefroid2008grammar, jones1999iogrammars, gorbunov2010autofuzz, aschermann2019nautilus}.

So, how about having a state model in which the individual message languages are specified by grammars?
Alas, grammars alone do not suffice either.
Looking again at \cref{fig:smtp-sequence}, we see that after the {\client \nonterm{client:HELO}} message \texttt{\client HELO relay.example.org}, the server replies with a {\server \nonterm{server:hello}} message \texttt{\server 250 Hello relay.example.org}---that is, it \emph{duplicates} the host name it received earlier.
Such equality cannot be expressed in a context-free grammar, let alone a state model; neither can other widely used
features such as checksums, length encodings and compression.

In practice, such \emph{semantic} properties are typically expressed in \emph{natural language}.
For instance, RFC documents, which document the protocols of the Internet, specify all semantic protocol properties in English.
(It may be surprising or even shocking, but the Internet entirely runs on protocols \emph{that are not fully formally specified.})
For symbolic verification purposes, this is less of a problem, as one can (and often must) abstract away details that are not relevant for the goal---the exact syntax of messages may be irrelevant for protocol properties.
But shouldn't we have tools or approaches that would be able to systematically \emph{test} whether a protocol implementation conforms to its specification?

The recent concept of \emph{language-based testing}~\cite{dominicACM} puts forward a solution, enriching grammars with \emph{constraints} as \emph{logical predicates over nonterminals.}
Such constraints allow us to express \emph{semantic properties} of messages.
For our \SMTP protocol, a constraint such as
\begin{equation} 
\client \nonterm{client:HELO}.\nonterm{hostname} \any = \server \nonterm{server:hello}.\nonterm{hostname} \any
\label{eq:hostname}
\end{equation}
could express that the hostname in the server response must be the same as previously received in the client message.
In a similar vein, one can express that one element is a checksum over another $\bigl(\nonterm{hash} = \textnormal{sha256}(\nonterm{message})\bigr)$, or that an element must be of a certain length $\bigl(\textnormal{uint16}(\nonterm{length}) = |\nonterm{payload}|\bigr)$.
When generating strings, the fuzzer must ensure that they \emph{satisfy} the constraints; this can be done using symbolic execution~\cite{steinhofel2022input} or evolutionary algorithms~\cite{zamudio2025fandango}.
So far, though, language-based testing has been limited to \emph{stateless} systems, producing specification-conforming inputs and/or checking outputs for conformance.

In this paper, we propose a novel, principled approach to protocol testing.
\emph{Language-based protocol testing} is an extension of language-based testing, again combining a grammar to specify the syntax of strings (messages) with \emph{constraints} over grammar elements to express semantic properties and relations.
The key element of language-based protocol testing, however, is the concept of an \emph{\IO grammar,} an extension of context-free grammars in which each message element is tagged with the \emph{communication party that is in charge of producing it:}

\begin{description}
\item[Messages.]
The special feature of an \IO grammar is that in the resulting string, each symbol is tagged with the \emph{party} that is in charge of sending the symbol.
In the \IO grammar, we indicate such tags by prefixing a nonterminal with the party name and a colon.
The sequence $\langle\textit{exchange}\rangle ::= \text{\client $\langle\textit{client:HELO}\rangle$}\ \text{\server $\langle\textit{server:hello}\rangle$}$ thus defines an exchange, consisting of a message {\text{\client $\langle\textit{client:HELO}\rangle$}} sent by the client, followed by a {\server \nonterm{server:hello}} response sent by the server.

\item[States and Transitions.] Since a state model represents a regular language as possible sequences of states, an \IO grammar can also embed a \emph{state model.}
This is done using a standard embedding technique:
\begin{enumerate}
  \item Every \emph{state} $S$ in the model becomes a nonterminal $S$ in the \IO grammar;
  \item Every \emph{labeled transition} $S_1 \to S_2$ labeled with messages~$M$ becomes an expansion of $S_1$~into~$S_2$ via $M$, or $S_1 ::= M \quad S_2$
  \item If multiple alternatives are outgoing from a state $S$, each of them becomes a separate alternative for the expansion of~$S$.
\end{enumerate}
The \IO grammar thus encodes not only the syntax of messages, but also the possible \emph{paths} through the state model.
Both states and messages are thus represented as nonterminals.
\end{description}

\begin{figure}
\renewcommand{\litleft}{`\bgroup\ulitleft\ttfamily\bfseries}

\begin{densegrammar}\normalsize
\any <start> ::= <connect>

\any <connect> ::= \server <server:id> \any <helo>

\server <id> ::= `220 ' <hostname> ` ESMTP Postfix\crlf'

\any <helo> ::= \client <client:HELO> \any $\bigl($\server{}<server:hello> \any <from> | \server <server:error>\any{}$\bigr)$

\client <HELO> ::= `HELO ' <hostname> `\crlf'

\server <hello> ::= `250 Hello ' <hostname> `, glad to meet you\crlf' \any <from>

\server <error> ::= `5' <digit> <digit> ` ' <error\_message> `\crlf'

\server <error\_message> ::= <message>

\any <from> ::= \client <client:MAIL_FROM> \any $\bigl($\server <server:ok> \any <to> | \server <server:error>\any$\bigr)$

\client <MAIL_FROM> ::= `MAIL FROM:<' <email> `>\crlf'

\server <ok> ::= `250 Ok\crlf'

\any <to> ::= \client <client:RCPT_TO>
    \any $\bigl($\server <server:ok> \any <data> | \server <server:ok> \any <to> | \server <server:error>\any{}$\bigr)$

\client <RCPT_TO> ::= `RCPT TO:<' <email> `>\crlf'

\any <data> ::= \client <client:DATA> \server <server:end_data> \client <client:message> 
     \any $\bigl($\server <server:ok> \any <quit> | \server <server:error>\any{}$\bigr)$

\client <DATA> ::= `DATA\crlf'

\server <end_data> ::= `354 End data with <CR><LF>.<CR><LF>\crlf'

\client <message> ::= \textbf{\ttfamily r}`[^.\crlf]*\crlf[.]\crlf'

\any <quit> ::= \client <client:QUIT> \server <server:bye>

\client <QUIT> ::= `QUIT\crlf'

\server <bye> ::= `221 Bye\crlf'

\end{densegrammar}
\caption{An \IO grammar for the \SMTP interaction in \Cref{fig:smtp-sequence}.
\nonterm{email} and
\nonterm{hostname} are defined externally;
\nonterm{message} is a regular expression.}
\label{fig:smtp-grammar}
\end{figure}

Let us have a look at a (reduced) \IO grammar for the \SMTP protocol, shown in \cref{fig:smtp-grammar}, derived from \cref{fig:smtp-lts} using the above rules.
\iffalse
At first, an \IO grammar is a \emph{context-free grammar,} consisting of \emph{rules} that expand \emph{nonterminals} into alternative sequences of terminal and nonterminal symbols.
The special feature of an \IO grammar is that in the resulting string, each symbol is tagged with the \emph{party} that is in charge of sending the symbol.
In the \IO grammar, we indicate such tags by prefixing a nonterminal with the party name and a colon.
The \nonterm{connect} nonterminal, for example, first expands into \nonterm{server:id}, indicating that the server sends the \nonterm{id} message.
It then expands into \nonterm{helo}, which expands into \nonterm{client:HELO}.
This is either followed by \nonterm{server:hello}, indicating that the server accepts the connection (followed by \nonterm{from} and more interactions); or by \nonterm{server:error}, producing an error message and ending the interaction.
\fi
This \IO grammar can be used as a complete specification for a \emph{client-side} test generator (filling in nonterminals such as \client \nonterm{hostname}\any, \client \nonterm{email} \any or \client \nonterm{message} \any with according values), but equally for a \emph{server-side} test generator (filling in nonterminals such as \server \nonterm{hostname} \any or \server \nonterm{error\_message} \any with according values).
A server-side test generator can also exercise the various points in the protocol where it can produce \server \nonterm{error} \any messages and thus systematically \emph{cover} the transitions in the state model shown in \cref{fig:smtp-lts} embedded in the \SMTP{} \IO grammar.
Furthermore, \emph{grammar coverage metrics} such as $k$-path~\cite{havrikov2019} naturally extend to \IO grammars, allowing to systematically guide generation towards yet uncovered grammar alternatives in input \emph{and} state space.

At the same time, \emph{constraints} over grammar elements serve to specify details of the input format (as in \Cref{eq:hostname}), but can also serve as \emph{oracles,} having the fuzzer \emph{checks} if the constraints are satisfied.
In our \SMTP example, the constraint
$
    \server \nonterm{error} \any \notin \nonterm{start}
$
expresses the set of interactions without server errors.
If the server issues an \nonterm{error} message, this violates the constraint, and thus, the interaction does not adhere to the protocol specification.

If constraints are Turing-complete, they can express any computable property; and in principle, one could thus express (or actually implement) any protocol as a single constraint.
This, however, would not be different from implementing a protocol-specific fuzzer and/or parser in code.
In practice, it is much easier to specify an \IO grammar for the \emph{syntax} of a protocol, and declarative \emph{constraints} for its (typically few) semantic properties.
To the best of our knowledge, this paper presents \emph{the first formal foundation that can be used for testing real-world protocols in all details.}

\iffalse
We have extended the \FANDANGO fuzzer~\cite{zamudio2025fandango} with support for \IO grammars.
\FANDANGO is a language-based fuzzer that uses a grammar to generate strings, which are then evolved to satisfy given constraints.
Constraints are specified as Python functions (including self-defined functions), making the constraint language very expressive.
\IIO-enabled \FANDANGO can act as any set of parties in the given protocol, allowing it to fuzz clients and servers alike.
\fi
Specifically, we make the following contributions:

\begin{description}
    \item[A formal specification language for protocol testing.] We introduce \emph{\IO grammars} for specifying messages, states, and transitions.
    As grammars, they open up an immense body of work for tasks such as
    \begin{enumerate*}[label=(\arabic*)]
     \item \emph{generating} strings from grammars~\cite{burkhardt1967,purdom1972};
     \item \emph{parsing} and checking strings against grammars~\cite{earley1970cacm};
     \item \emph{composing} and transforming grammars~\cite{aho1972theory};
     \item embedding \emph{state models} (and all their related work, such as model checking or program verification) into grammars~\cite{hopcroft2001introduction, utting2010practical, alur2004visibly};
     and
     \item measuring and establishing grammar \emph{coverage}~\cite{havrikov2019}.
    \end{enumerate*}

    \item[A single specification for multiple parties.] One \IO grammar specifies the behavior of all parties in a protocol, whether client or server.
    A test generator can thus act on the client side or on the server side.
    To the best of our knowledge, our approach is the first to specify (and instantiate) interactions for all parties during protocol testing.

    \item[Declarative specifications of semantic protocol properties.] We have extended the \FANDANGO fuzzer~\cite{zamudio2025fandango} with support for \IO grammars.
    This allows us to specify and solve \emph{constraints} as predicates over grammar elements (extended for \IO grammars), formalizing semantic properties of protocols in a declarative way.
    \item[Systematic coverage of state and message space.]
    Building on \FANDANGO also allows us to achieve \emph{coverage} in state and message space.
    We adapt the concept of \emph{$k$-path coverage}~\cite{havrikov2019} to \IO grammars including non-controlled parties, and can thus systematically explore and cover all production alternatives, including messages, responses, (induced) states and transitions, in a unified fashion,
    thereby also increasing code coverage in the system under test (SUT).
    \item[A universal way to specify \emph{oracles} in protocols.] By parsing outputs according to the grammar, we decompose responses into individual elements; using constraints over elements (including earlier interactions), we can check them for correctness.
    Accessing elements also allows \emph{reacting} to responses such as opening a port listed in an \FTP server response.
   
    \iffalse 
    \item[A specification for mocking and monitoring.] A rich combination of grammars and constraints allows tools to \emph{generate} responses for given messages, effectively acting as a mock object.
    \emph{Parsing} existing interactions between parties yields a \emph{monitor} that can check whether the interaction adheres to the protocol or given constraints.
    \fi
\end{description}

\iffalse 
We have extended the \FANDANGO fuzzer~\cite{zamudio2025fandango} with support for \IO grammars.
\FANDANGO is a language-based fuzzer that uses a grammar to generate strings, which are then evolved to satisfy given constraints.
Constraints are specified as Python functions (including self-defined functions), making the constraint language Turing-complete.
Our prototype extends \FANDANGO with support for \IO grammars, allowing it to act as any set of parties in the given protocol.
\fi

The remainder of the paper is structured as follows.
The main contribution comes in \Cref{sec:approach} with a formal foundation on how to test protocols with \IO grammars, followed by \Cref{sec:impl}, describing our implementation on top of \FANDANGO.
\Cref{sec:casestudy} shows a set of case studies on how to use \FANDANGO with \IO grammars to test five different stateful implementations.
In \Cref{sec:evaluation}, we evaluate \FANDANGO with \IO grammars,
finding that
its specifications are concise (\Cref{sec:rq1-conciseness});
it has a high throughput (\Cref{sec:rq2-performance});
it quickly covers the entire \IO grammar (\Cref{sec:rq3-coverage});
its grammar coverage guidance is essential (\Cref{sec:rq4-guidance}); and
it can outperform mutation-based fuzzers such as \AFLNET (\Cref{sec:rq5-aflnet}) in terms of code coverage and bugs found.
After discussing related work in \Cref{sec:related-work}, \Cref{sec:conclusion} closes with conclusions and future work.

\FANDANGO and all experimental data are available as open source; see
\Cref{sec:conclusion} for details.

\section{Language-Based Protocol Testing}
\label{sec:approach}

\begin{figure}[t]
    \centering
    \includegraphics[width=\linewidth]{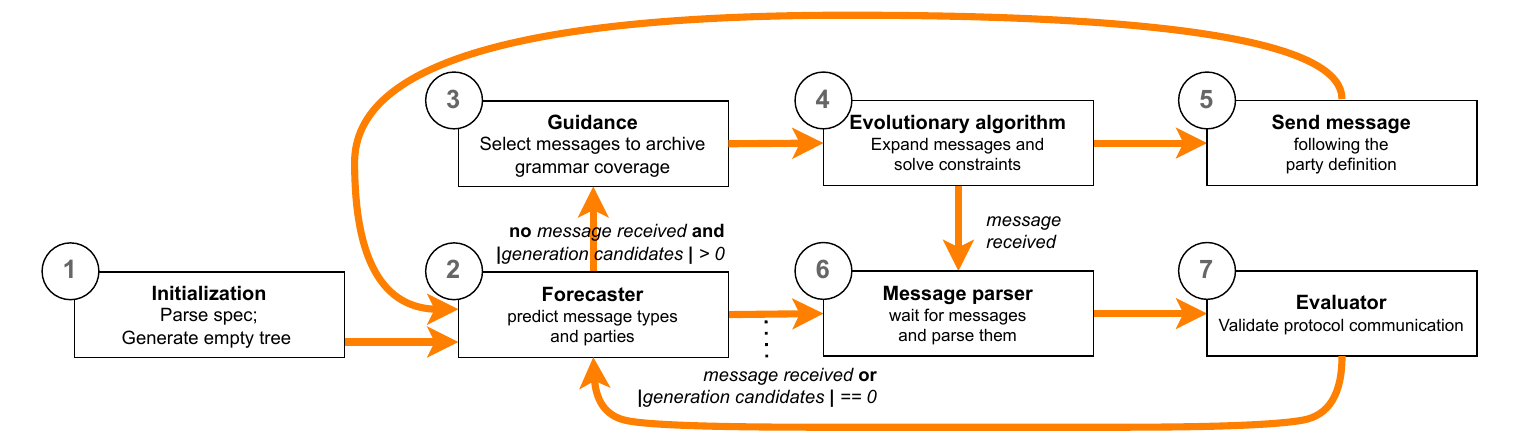}
    \caption[Processing \IO grammars]{Processing \IO grammars.
    After parsing an \IO grammar specification, \FANDANGO~\circled{1} initializes party definitions and generates an empty derivation tree.
    The \textit{forecaster}~\circled{2} predicts possible upcoming protocol message types.
    If these include messages from fuzzer-controlled parties and no external messages have been received, our coverage guidance~\circled{3} selects message types for generation~\circled{4}.
    If no external message is received, \circled{5} a generated message is transmitted.
    Otherwise, if an external message is received, \circled{6} it is parsed immediately.
    Finally, \circled{7} the resulting derivation tree is validated.
    The process repeats until the forecaster predicts no further messages.}
    \label{fig:flowchart}
\end{figure}

To test protocols with \IO grammars, we combine concepts from evolutionary algorithms~\cite{bartz2014evolutionary} with language-based software testing~\cite{dominicACM} to systematically explore the input and state space of \IO grammars.
\textit{Language-based Protocol Testing}, as we have named it, leverages the syntactical structure of interactions defined by \IO grammars and applies evolutionary operators to generate syntactically valid and semantically meaningful inputs, maximizing for structure diversity~\cite{havrikov2019}.
\IIO grammars provide a format that defines the message syntax and semantics as well as the communication model within a single specification, and the same \IO grammar can be used to test any party that takes part in the protocol. \Cref{fig:flowchart} visualizes how the components introduced in this chapter work together within an implementation for \IO grammars that we built on top of the \FANDANGO framework.

\subsection{Syntax and Semantics}

\begin{definition}[Grammar]
\label{def:grammar}
We start with the standard definition of a context-free grammar~$G$ as a tuple $G = (\Sigma, V, S, R)$, comprising a set of terminal symbols $\Sigma$, a set of nonterminal symbols~$V$, a start symbol $S \in V$, and a set of production rules $R$. Each rule is of the form $A \rightarrow \alpha$, where $A \in V$ is a nonterminal and $\alpha \in (V \cup \Sigma)^*$ is a sequence of terminals and/or nonterminals~\cite{zamudio2025fandango}.
\end{definition}

\begin{definition}[\IIO Grammar]
\label{def:io-grammar}
We define an \emph{\IO grammar} as a tuple $\IG = (G, \mathcal{I}, \pi)$, extending the standard grammar $G$. Here, $\mathcal{I}$ is a finite set of \emph{parties} (e.g., $\{\text{Client}, \text{Server}\}$), and $\pi$ is a mapping that assigns communication roles to nonterminals:
\[
\pi: V \to (\mathcal{I} \cup \{\varepsilon\}) \times (\mathcal{I} \cup \{\varepsilon\})
\]
For any nonterminal $A \in V$, the mapping $\pi(A) = (I_s, I_r)$ defines a \emph{sender} $I_s$ and a \emph{receiver} $I_r$, subject to the following conditions:
\begin{itemize}
    \item If $I_s \ne \varepsilon$, $A$ is a \emph{message nonterminal}, denoting that the subtree derived from $A$ constitutes a discrete message sent by $I_s$.
    \item If $I_r = \varepsilon$, the receiver is undefined at this level and must be inferred from the derivation context.
    \item We denote message nonterminals using the notation \nonterm{sender:receiver:name} or \nonterm{sender:name}.
    \item The combination $I_s = \varepsilon$ and $I_r \ne \varepsilon$ is not permitted. A nonterminal cannot have a defined receiver if it lacks a defined sender (i.e., purely structural nonterminals cannot be "received").
\end{itemize}
The production rules $R$ remain standard context-free rules, defining the structure of the messages and the protocol flow.
\end{definition}

An \IO grammar is processed depth-first, from left to right; consequently, messages are generated sequentially. This enables the test generator to perform a \emph{context switch} between message generation and parsing after each message is completed.
For example, in \cref{fig:smtp-grammar}, the nonterminal \nonterm{connect} (with $\pi(\text{\nonterm{connect}}) = (\varepsilon, \varepsilon)$) expands into \nonterm{server:id}\nonterm{helo}.
The element \nonterm{server:id} is a message nonterminal where the sender is the \textit{server}.
Such nonterminals are fully expanded to form a complete message subtree before transmission.
 
\begin{definition}[Constrained \IIO Grammar]
Building upon \cref{def:io-grammar}, a \emph{constrained \IO grammar} is a tuple $G_C = (\IG, \Phi)$, where $\Phi$ is a finite set of \emph{constraints}.
Each constraint $\phi \in \Phi$ is a logical predicate $\phi: \mathcal{D}_G \to \{\text{true}, \text{false}\}$ defined over the set of valid derivation trees $\mathcal{D}_G$.
Constraints restrict the valid input space by enforcing semantic properties (e.g., field lengths, checksums, or value dependencies) that cannot be expressed in pure context-free form.
\end{definition}

In this paper, we assume all \IO grammars are constrained ($\Phi \neq \emptyset$), as valid protocol communications inherently rely on semantic conditions.

\subsection{Instantiation}
\label{sec:instantiation}

\begin{wrapfigure}[12]{R}{0.6\linewidth}
    \centering
    \includegraphics[width=\linewidth]{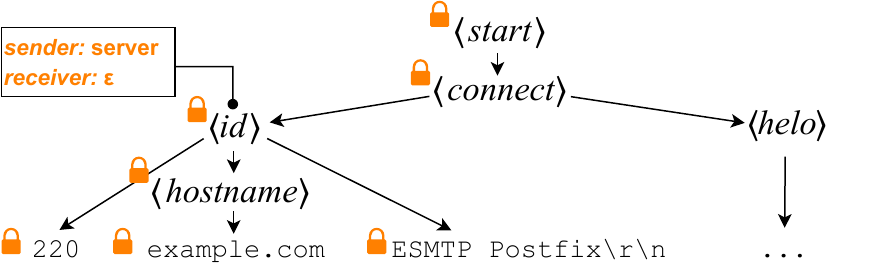}
    \caption{Partial derivation tree produced by the grammar in \Cref{fig:smtp-grammar}.
    Each node inherits the \emph{sender} and \emph{receiver} from $\pi(A)$, and maintains a mutable \emph{read\_only} flag ({\scriptsize \faLock}).}
    \label{fig:example-devtree}
\end{wrapfigure}

To represent the structure of an input during fuzzing, we use \textit{derivation trees}.
A \textit{derivation tree} \( T \) is an ordered tree that encodes the syntactic structure of a string according to the grammar $G$ (\cref{def:grammar}).
Each internal node \( n \) in \( T \) is labeled with a nonterminal \( A \in V \) and expands according to a production rule in \( R \); leaf nodes correspond to terminals in \( \Sigma \) or the empty string \( \varepsilon \).
The yield of the tree forms a string in \( \Sigma^* \) generated by \( G \)~\cite{zamudio2025fandango}.

In our approach, we extend the standard derivation tree to support protocol state tracking.
Every node \( n \) labeled with a nonterminal \( A \) inherits the communication roles defined by the mapping $\pi(A) = (I_s, I_r)$ from \cref{def:io-grammar}.
This explicitly associates message boundaries and direction (sender/receiver) with the tree structure.
Additionally, we introduce a boolean \emph{read-only} flag for each node.
When enabled, this flag prevents \FANDANGO from mutating the node (e.g., to satisfy a constraint).
\Cref{fig:example-devtree} illustrates this augmented tree structure.

While the grammar defines the abstract set of parties $\mathcal{I}$, the fuzzer requires concrete implementations to handle network transmission.
The \IO grammar specification includes bindings that define how these parties interact.
This design allows users to implement custom communication logic, making the specification highly versatile.
A typical party implementation defines three core methods, as shown in \Cref{fig:party-spec-gpt}:

\begin{figure}[h]
    \centering
   
    \begin{minipage}[t]{0.45\linewidth}
        \vspace{0pt}
        \begin{minted}[fontsize=\footnotesize, frame=single, framesep=5pt]{python}
class SMTPClient(NetworkParty):
    def __init__(self):
        super().__init__(
            "tcp://127.0.0.1:25",
            connection_mode=ConnectionMode.CONNECT
        )
        self.start()

    def send(self, message, target):
        super().send(message, target)

    def receive(self, message, sender):
        super().receive(message, sender)
        \end{minted}
    \end{minipage}
    \hfill
    \begin{minipage}[t]{0.5\linewidth}
        \vspace{0pt}
         \begin{itemize}
        
            \item \texttt{\_\_init\_\_()}: Initializes the communication context and determines if the party is simulated by the fuzzer or represents the external SUT, as well as the endpoint address.
        
            \item \texttt{send()}: Gives the user the ability to perform additional transformations to the message before transmission.

            \item \texttt{receive():} Gives the user the ability to perform additional transformations to the message after reception.
        
        \end{itemize} 
    \end{minipage}
    \caption{Snippet of a sample definition of an SMTP party.}
    \label{fig:party-spec-gpt}
\end{figure}

The \FANDANGO command line interface comes with predefined party implementations for common client and server scenarios.
To test an \SMTP server, for instance, it is not necessary to explicitly define a client.
Instead, it suffices to invoke \FANDANGO with the option \texttt{-{}-client tcp://localhost:8025}, and the given \SMTP \IO grammar; it then connects as client to port~25 on the local host using tcp.

\subsection{Forecaster}
\label{sec:forecaster}

The grammar~$G$ defines the admissible execution flow of the protocol, dictating which party~$I$ is allowed to send which message~$M$ at any given point.
While regular grammars correspond to finite state machines, the context-free grammars used here define a \emph{pushdown automaton}, where the validation state includes a stack of open nonterminals.
Consequently, the subset of admissible messages~$M$ changes dynamically with each newly sent or received message.
Furthermore, a grammar~$G$ may contain $\varepsilon$-transitions or multiple rules for the same nonterminal, rendering the protocol state nondeterministic.

This creates a challenge where multiple valid \emph{partial derivation trees} may exist for the same observed sequence of messages.
We define a \emph{partial derivation tree}~$T$ as a tree that matches $G$ starting from~$S$, but where the yield is only a prefix of a complete, valid string.
Each possible partial tree~$T$ represents a distinct hypothesis about the current state of the protocol, potentially allowing different sets of future messages.

To derive all possible partial trees~$T$ consistent with the observed traffic, we use a modified \emph{Earley parser}~\cite{earley1970cacm}.
Since we assume \emph{leftmost derivations} (processing the input from left to right), the next symbol to be generated or parsed always corresponds to the first unexpanded nonterminal in the tree's next step.
The forecasting mechanism inspects these unexpanded nonterminals to identify the set of valid next steps.
Based on the grammar's structure, it determines the set of nonterminals~$A$ that can be expanded next, ensuring that the resulting tree~$T$ remains consistent with~$G$.
The output is a set of \emph{predictions}, where each prediction is a 5-tuple:
\[
    P = (T, n, I_s, I_r, A)
\]
Here:
\begin{enumerate}
    \item $T$ is the partial derivation tree representing the current protocol state;
    \item $n$ is the reference to the leaf node in~$T$ (the 'hook-in' point) where the expansion will occur;
    \item $A$ is the nonterminal labeling node~$n$ (where $\pi(n) = (I_s, I_r)$);
    \item $I_s$ is the sending party derived from $\pi(A)$; and
    \item $I_r$ is the receiving party derived from $\pi(A)$.
\end{enumerate}
This tuple provides the fuzzer with the precise context needed to generate or parse the next message.

\subsection{Responsibility Manager}

The \IO grammar $\IG$ defines the admissible sequences of messages exchanged between the set of participating parties $\mathcal{I}$.
Crucially, the fuzzing engine must distinguish between its own actions and those of the system under test.
We formally partition the set of parties $\mathcal{I}$ into two disjoint subsets:
\begin{itemize}
    \item $\mathcal{I}_{\text{fuzz}} \subset \mathcal{I}$: Parties simulated and controlled by the fuzzer.
    \item $\mathcal{I}_{\text{ext}} \subset \mathcal{I}$: External parties (the system under test) whose messages must be observed and parsed.
\end{itemize}

As described in \cref{sec:forecaster}, the Forecaster outputs a set of valid next steps, where each prediction includes the sending party $I_s$.
The Responsibility Manager uses these predictions to resolve the \emph{send-receive dilemma}: determining whether the fuzzer should actively generate a message or wait for input.
The decision logic proceeds as follows:
\begin{enumerate}
    \item \textbf{Analyze Predictions:} The manager inspects the set of predicted next steps. If the set contains any prediction where the sender $I_s \in \mathcal{I}_{\text{fuzz}}$, the fuzzer prepares to generate a message.
    \item \textbf{Check Input Queue:} Before generation begins, the manager checks if any external messages have already been received from a party $I \in \mathcal{I}_{\text{ext}}$.
    \item \textbf{Priority Handling:}
    \begin{itemize}
        \item If an external message is pending, it takes precedence. The fuzzer immediately routes the message to the parser to update the protocol state.
        \item If no external message is pending and the fuzzer is allowed to send ($I_s \in \mathcal{I}_{\text{fuzz}}$), the generation process begins.
    \end{itemize}
    \item \textbf{Race Condition Mitigation:} Even after a message candidate is generated, the fuzzer performs a final check of the input queue before transmission. If an external message arrived during the generation phase, the generated candidate is discarded, and the incoming message is processed instead.
\end{enumerate}

This "listen-before-talk" approach ensures that the fuzzer remains synchronized with the external system, prioritizing the processing of state changes initiated by the SUT over its own generated traffic.

\subsection{Message Parsing}
\label{sec:message-parsing}

In network protocols, a distinct logical message~$M$ is often split into multiple transmission units or \emph{fragments} $F_1, \ldots, F_n$ due to transport layer constraints (e.g., TCP segmentation~\cite{postel1983tcp}).
To reconstruct the derivation tree~$T$, the parser must reassemble these fragments into a contiguous stream before matching them against the grammar.
We assume strictly ordered delivery, where fragments $F_i$ arrive sequentially and are not interleaved with unrelated data from the same party.
Our system maintains a \emph{cumulative buffer} $B_k$ representing the concatenation of received fragments:
\[
    B_k = F_1 \circ F_2 \circ \dots \circ F_k
\]
where $\circ$ denotes sequential concatenation.
Upon receiving a new fragment $F_k$, the parser attempts to construct a valid derivation tree using the set of \emph{expected nonterminals} $\mathcal{A}_{exp}$ provided by the Forecaster's predictions (see \cref{sec:forecaster}).
For each candidate nonterminal $A \in \mathcal{A}_{exp}$ and the current buffer $B_k$, the Earley parser determines the state of the parse:
\begin{enumerate}
    \item \textbf{Complete Match:} The buffer $B_k$ forms a complete string in $L(A)$. A valid derivation tree $T_A$ exists.
    \item \textbf{Valid Prefix:} The buffer $B_k$ is not yet a complete message but corresponds to the beginning of a valid string in $L(A)$. In this case, the system waits for further fragments.
    \item \textbf{Mismatch:} The buffer $B_k$ cannot be derived from $A$ (neither as a complete string nor as a prefix). The nonterminal $A$ is discarded from the candidate set.
\end{enumerate}

\subsubsection{Ambiguity and Greedy Parsing}
A common challenge arises when $B_k$ is a complete match for $A$, but could also be a valid prefix for a longer message (e.g., a command followed by optional arguments).
To address this, we employ a \emph{greedy strategy with timeout}:
If a complete parse is found but the buffer remains a valid prefix for an extended derivation, the parser tentatively caches the result and waits for a short duration $\Delta t$.
\begin{itemize}
    \item If a subsequent fragment $F_{k+1}$ arrives and extends the validity, the parser proceeds with $B_{k+1}$.
    \item If the timeout expires or $F_{k+1}$ renders the buffer invalid, the cached complete parse is committed and returned.
\end{itemize}

If the candidate set $\mathcal{A}_{exp}$ becomes empty (no expected nonterminal matches the buffer), the parsing process aborts, treating the input as a protocol violation.

\subsection{Solving Constraints}
\label{sec:solving-constraints}

To systematically produce valid and diverse protocol communications, it is necessary to explore the constrained input space defined by $G_C = (\IG, \Phi)$.
Since the constraint set $\Phi$ often defines a non-linear and discontinuous search landscape, simple random generation is insufficient.
We therefore adopt a \emph{Search-Based Software Testing}~\cite{mcminn2011search} (SBST) approach, specifically leveraging an evolutionary algorithm to iteratively refine a population of derivation trees.

We extend the genetic operators of the \FANDANGO generator~\cite{zamudio2025fandango} to support the party-annotated derivation trees defined in this work.
The goal is to evolve trees $T$ that maximize the satisfaction of semantic constraints $\Phi$ while preserving syntactic correctness with respect to $\IG$.
The evolutionary process proceeds as follows:

\begin{enumerate}
    \item \textbf{Evaluation:} In each generation, we calculate a \emph{fitness score} for every candidate tree $T$. The fitness function quantifies the degree to which $T$ satisfies the constraints in $\Phi$ (e.g., matching length fields or checksums).
    \item \textbf{Selection:} Based on these fitness scores, we select the best-performing candidates to serve as parents for the next generation.
    \item \textbf{Variation:} We apply grammar-aware variation operators to generate offspring:
    \begin{enumerate}
        \item \emph{Crossover:} Recombines subtrees from two parent derivation trees. This facilitates the exchange of valid structural features (e.g., a correct header sequence) between candidates.
        \item \emph{Mutation:} Introduces stochastic changes to specific subtrees. This operator ensures diversity and prevents the search from converging prematurely to local optima.
    \end{enumerate}
    \item \textbf{Termination:} The process repeats until a candidate $T$ fully satisfying $\Phi$ is found, or a computational budget is exhausted.
\end{enumerate}

Crucially, these variation operators are designed to be \emph{closed} under the grammar $\IG$, meaning they always produce syntactically valid derivation trees.
This allows the search to focus entirely on satisfying the semantic properties defined by $\Phi$, significantly increasing the probability of generating valid, high-coverage protocol interactions.

\subsection{Algorithm Workflow}
\label{sec:algorithm-workflow}

\begin{wrapfigure}[18]{R}{0.52\textwidth} 
    \vspace{-2\baselineskip}
    \begin{minipage}{0.52\textwidth}
    \begin{algorithm}[H]
    \caption{Protocol Message Exchange}
    \label{alg:fandangoio}
    \footnotesize
    \begin{algorithmic}[1]
    \State $T \gets \textit{init\_tree}()$
    \State $\textit{io} \gets \textit{IoInterface}()$

    \While{\textbf{true}}
       
        \State $P \gets \textit{predict\_messages}(T)$
        
        \If{$\textit{is\_empty}(P)$} \Comment{Session Complete}
            \State \Return $T$
        \EndIf

        \State $P_{\textit{fuzz}} \gets \{ p \in P \mid \textit{sender}(p) \in \mathcal{I}_{\textit{fuzz}} \}$
        \State $P_{\textit{ext}} \gets \{ p \in P \mid \textit{sender}(p) \in \mathcal{I}_{\textit{ext}} \}$

        \If{$\textit{io.has\_data}() \lor \textit{is\_empty}(P_{\textit{fuzz}})$}
           
            \State $\textit{fragments} \gets \textit{io.wait\_data}()$
            \State $T \gets \textit{parse}(\textit{fragments}, P_{\textit{ext}})$
        \Else
            \State $P_{\textit{fuzz}} \gets \textit{select\_prediction}(P_{\textit{fuzz}})$
           
            \State $M \gets \textit{solve\_constraints}(P_{\textit{fuzz}})$

            \If{$\textit{io.has\_data}()$}
                 \State \textbf{continue} \Comment{Abort}
            \EndIf
            
            \State $\textit{io.transmit}(M)$
            \State $T.\textit{append}(M)$
        \EndIf
        
        \State $T.\textit{set\_read\_only}()$
    \EndWhile
    \end{algorithmic}
    \end{algorithm}
    \end{minipage}
\end{wrapfigure}

The overall execution of a single protocol session is formalized in \cref{alg:fandangoio}.
Given an \IO grammar $\IG$, the algorithm iteratively constructs a derivation tree $T$ representing the communication trace.

The process begins by initializing an empty derivation tree.
In each iteration of the main loop, the \emph{Forecaster} (see \cref{sec:forecaster}) analyzes the current frontier of $T$ to predict the set of admissible next messages.
Based on these predictions, the \emph{Responsibility Manager} determines the next action:
\begin{itemize}
    \item \textbf{Reception:} If an external message is received (or prioritized), the system parses the input using the \emph{Earley parser} to extend $T$.
    \item \textbf{Generation:} If the fuzzer is responsible for the next step, we select the next prediction according to our guidance metric (\Cref{sec:impl}), and run the evolutionary constraint solver (see \cref{sec:solving-constraints}) to generate a valid message derivation, which is then serialized and transmitted.
\end{itemize}

The derivation tree $T$ is incrementally updated with each new message. The algorithm terminates when the derivation is complete (i.e., the tree contains no unexpanded nonterminals) or when the Forecaster predicts no valid future transitions, indicating the end of the protocol session.
Finally, the complete derivation tree $T$, encoding the full interaction trace, is returned for analysis.

\section{Implementation}
\label{sec:impl}

We have implemented the approach formally defined in \cref{sec:approach} as an extension of the \FANDANGO fuzzer~\cite{zamudio2025fandango}.
\FANDANGO is a modular, language-based fuzzing framework that facilitates the generation of inputs satisfying both syntactic rules (via grammars) and semantic predicates (via constraints).
Our extension integrates the \IO grammar formalism (\IG), the state-tracking derivation tree structure, and the Python-based party interface described in \cref{sec:approach}.

While \cref{alg:fandangoio} (\cref{sec:algorithm-workflow}) defines the fundamental message exchange loop, the efficacy of the fuzzer depends on its ability to explore the protocol state space systematically.
To this end, our implementation augments the core loop with a \textbf{coverage-driven guidance system}.
This system aims to achieve $k$-path coverage by coordinating three specialized components: a \textit{Target Selector}, a \textit{Navigator}, and a \textit{Runtime Message Guider}.
The interaction and logic of these components are detailed in the following subsections.

\subsection{Covering Production Alternatives, States, and Transitions}
\label{sec:k-path}

Comprehensive protocol testing requires covering the specification as thoroughly as possible.
In the context of grammar-based fuzzing, Havrikov et al.~\cite{havrikov2019} introduced the concept of \emph{$k$-path coverage}, which ensures that a producer explores all possible combinations of production alternatives up to a depth of $k$ in the derivation tree.
Coverage at $k=1$ implies that every alternative in the grammar is exercised at least once; $k=2$ ensures that all consecutive pairs of alternatives are covered.

Since an \IO grammar implicitly encodes the protocol's finite state machine, the grammar structure reflects the state model (see \Cref{fig:smtp-grammar}).
Consequently, a set of interactions achieving $k$-path coverage ($k=1$) effectively \emph{covers all transitions in the embedded state model.}
For the \SMTP example, this guarantees that for every command, both the successful execution (leading to the next state) and the error path (leading to a rejection) are tested.

\begin{wrapfigure}[17]{R}{0.5\textwidth}
    \vspace{-2\baselineskip}
    \begin{minipage}{0.5\textwidth}
    \begin{algorithm}[H]
    \caption{Target Selection Strategy}
    \footnotesize
    \label{alg:targetselection}
    \begin{algorithmic}[1]
    \Require Grammar $G$, Corpus $\mathcal{T} = \{T_1, \dots, T_n\}$, depth $k$
    \Ensure A target path $\tau$ to explore

    \State $\mathcal{K}_{all} \gets \textit{AllPaths}(G, k)$
    \State $\mathcal{K}_{cov} \gets \bigcup_{T \in \mathcal{T}} \textit{ExtractPaths}(T, k)$
    \State $\mathcal{K}_{uncov} \gets \mathcal{K}_{all} \setminus \mathcal{K}_{cov}$

    \State $\mathcal{P}_{target} \gets \emptyset$
    \ForAll{$p \in \mathcal{K}_{uncov}$}
        \State $p' \gets \textit{project\_states}(p)$
        \If{$|p'| > 0$}
            \State $\mathcal{P}_{target} \gets \mathcal{P}_{target} \cup \{p'\}$
        \EndIf
    \EndFor

    \If{$\mathcal{P}_{target} \neq \emptyset$}
       
        \State $\textit{Schedule} \gets \textit{StatePowerSchedule}()$
        \State $\tau \gets \textit{Schedule.sample}(\mathcal{P}_{target})$
    \Else
       
        \State $\mathcal{C}_{msg} \gets \textit{MsgTypeCoverage}(G, \mathcal{T})$
        \State $\textit{Schedule} \gets \textit{MsgPowerSchedule}()$
        \State $\tau \gets \textit{Schedule.sample}(\mathcal{C}_{msg})$
    \EndIf

    \State \Return $\tau$
    \end{algorithmic}
    \end{algorithm}
    \end{minipage}
\end{wrapfigure}

However, the ability to \emph{measure} coverage does not imply the ability to \emph{enforce} it.
While achieving $k$-path coverage is a solved problem for standard inputs~\cite{havrikov2019}, our \emph{interactions} involve an external system under test whose responses are outside our direct control.
Therefore, we require a production strategy that \emph{detects} coverage gaps and actively \emph{guides} the interaction towards these missing states.

To achieve full $k$-path coverage for an \IO grammar, we employ three distinct components (see \Cref{alg:targetselection} for the selection logic):
\begin{enumerate*}[label=(\arabic*)]
    \item A \textbf{Target Selection Algorithm} that identifies all uncovered $k$-paths and selects a specific path to target;
    \item A \textbf{Navigator} that computes the sequence of messages and states required to reach the selected $k$-path; and
    \item A \textbf{Message Guider} that monitors the current session, selecting messages that adhere to the navigator's plan and aborting the run if the protocol state diverges.
\end{enumerate*}

\subsubsection{Target Selection and Power Schedules}
\label{sec:power-schedule}

The \textbf{Target Selection Algorithm}, detailed in \cref{alg:targetselection}, is responsible for identifying structural gaps in the current testing corpus and selecting an optimal trajectory for the next fuzzing iteration.
First, the algorithm computes the set of all valid $k$-paths $\mathcal{K}_G$ inherent to the grammar $G$.
It then identifies the set of \emph{uncovered} paths $\mathcal{K}_U$ by subtracting the set of paths already present in the current population of derivation trees $\mathcal{T} = \{T_1, \dots, T_n\}$:
\[
    \mathcal{K}_U = \mathcal{K}_G \setminus \bigcup_{T \in \mathcal{T}} \text{paths}_k(T)
\]

\vspace{10pt}

To distinguish between high-level protocol state transitions and low-level payload structures, each path $p \in \mathcal{K}_U$ is projected onto the \emph{state sub-grammar}.
We define this projection $\Pi_{\text{state}}(p)$ as the subsequence of $p$ containing only nonterminals reachable from the start symbol $S$, excluding those belonging to message-payload definitions.
This yields the set of uncovered state paths $\mathcal{K}_{U,S}$.
The selection logic proceeds in two phases based on the emptiness of $\mathcal{K}_{U,S}$:

\paragraph{Phase 1: State Exploration.}
If $\mathcal{K}_{U,S} \neq \emptyset$, the fuzzer prioritizes exploring new protocol states.
A target path $p \in \mathcal{K}_{U,S}$ is selected using a \emph{power schedule} that assigns an energy value $E(p)$ to each candidate.

\noindent
\begin{minipage}{\linewidth}
\vspace{6pt}
\begin{equation}
    E(p) := \frac{\text{length}(p)}{\text{freq}(p) + 1}
\end{equation}
\vspace{6pt}
\end{minipage}

\noindent where $\text{length}(p)$ is the number of nodes in the path, and $\text{freq}(p)$ denotes the number of times $p$ has been previously selected as a target.
This schedule maximizes coverage efficiency by prioritizing:
\begin{enumerate*}[label=(\roman*)]
    \item \emph{Deep paths} (higher length), pushing the protocol into deeper states; and
    \item \emph{Neglected paths} (lower frequency), ensuring uniform exploration of the state space.
\end{enumerate*}

\paragraph{Phase 2: Message Diversity.}
If $\mathcal{K}_{U,S} = \emptyset$, the protocol's state model is considered fully covered.
The algorithm shifts focus to maximizing the internal diversity of individual messages.
For each message type $M$, a coverage score $C(M) \in [0, 1]$ is computed based on the saturation of its internal production rules in $\mathcal{T}$.
The energy assignment becomes:

\noindent
\begin{minipage}{\linewidth}
\vspace{6pt}
\begin{equation}
    E(M) := \frac{C(M)}{\text{freq}(M) + 1}
\end{equation}
\vspace{6pt}
\end{minipage}

\noindent This guides the fuzzer to target message types that structurally allow for high variance but have been undersampled in previous iterations.

\begin{wrapfigure}[20]{R}{0.5\textwidth}
    \vspace{-1\baselineskip}
    \begin{minipage}{0.5\textwidth}
    \begin{algorithm}[H]
    \caption{Message Guidance Logic}
    \footnotesize
    \label{alg:messageguiding}
    \begin{algorithmic}[1]
    \Require Derivation Tree $T$, Predictions $P$
    \State \textbf{Global:} Target $\tau$, Plan $\pi$, Preserved $\mathcal{K}_{\text{session}}$

    \If{$\textit{is\_new\_session}(T)$}
       \State $\mathcal{K}_{\text{session}} \gets \emptyset$; $\pi \gets \emptyset$; $\tau \gets \bot$
    \EndIf

    \State $\pi \gets \textit{advance\_plan}(\pi, T)$

    \If{$\textit{is\_empty}(\pi)$}
       
        \If{$\tau \neq \bot \land \textit{covers}(T, \tau)$}
            \State $\mathcal{K}_{\text{session}} \gets \mathcal{K}_{\text{session}} \cup \{\tau\}$
        \EndIf

        \State $\tau \gets \textit{SelectTarget}(G, \mathcal{T} \cup \{T\})$
        
        \If{$\tau = \bot$} \Comment{All covered}
            \State $\pi \gets \textit{NavigateToEnd}(T)$
            \State $guide\_to\_end \gets \textbf{true}$
        \Else
           
            \State $\pi \gets \textit{Navigator}(T, \tau, \mathcal{K}_{\text{session}})$
        \EndIf
    \EndIf

    \State $P_{\textit{guided}} \gets \{ m \in P \mid m \text{ matches } \textit{head}(\pi) \}$

    \If{$P_{\textit{guided}} \neq \emptyset$}
        \State \Return $P_{\textit{guided}}$
    \Else
        \State \Return $P$ \Comment{Fallback: Divergence allowed}
    \EndIf
    \end{algorithmic}
    \end{algorithm}
    \end{minipage}
\end{wrapfigure}

\paragraph{Local Optimization.}
During the generation of a chosen target, the fuzzer performs local look-ahead checks.
When expanding a nonterminal, it inspects the grammar for immediate expansions that would cover remaining entries in $\mathcal{K}_U$.
If a covering expansion exists, it is prioritized.
If no such expansion is found after a configurable number of attempts, the fuzzer falls back to random selection to prevent stagnation.

\subsubsection{Navigation}
\label{sec:navigation}

The \textbf{Navigator} component is responsible for computing a concrete execution path from the current protocol state to a selected target $k$-path $\tau$.
We model the search space as a directed graph where nodes represent partial derivation trees and edges represent valid message expansions defined by the grammar $G$.
Since this space is infinite, the Navigator employs an $A^*$ search algorithm~\cite{a-star} with lazy expansion: nodes are instantiated only when visited, ensuring scalability.

The heuristic function for $A^*$ estimates the distance from the current nonterminal frontier to the target path $\tau$.
The search operates over complete $k$-paths rather than isolated symbols, ensuring that the computed path accounts for required context (e.g., prerequisite states).
If the target $\tau$ is unreachable from the current state (e.g., due to a protocol restriction), the Navigator computes a \emph{reset path} that terminates the current session and initiates a new one, ensuring that all targets in $\mathcal{K}_U$ remain eventually reachable.

\subsubsection{Message Guidance}
\label{sec:message-guidance}

The \textbf{Message Guider} acts as the runtime controller. It enforces the plan computed by the Navigator by filtering the predictions provided by the Forecaster.
Its primary role is to ensure that the fuzzer systematically covers the selected target $\tau$ while preserving any $k$-paths $\mathcal{K}_{\text{session}}$ already achieved in the current run.

The guidance logic is formalized in \cref{alg:messageguiding}.
The algorithm maintains a \emph{guidance plan} $\pi$, which is a sequence of expected derivation steps.
In each iteration, the Guider performs the following:
\begin{enumerate}
    \item \textbf{Synchronization:} It updates $\pi$ by consuming steps that match the messages observed since the last iteration. If the session has diverged from $\pi$ (or if $\pi$ is exhausted), the plan is considered invalid.
    \item \textbf{Target Verification:} If the plan is finished, the Guider checks if the target $\tau$ was successfully covered. If so, $\tau$ is added to the \emph{preserved set} $\mathcal{K}_{\text{session}}$. This set is passed to the Navigator to ensure future plans do not inadvertently violate constraints required to keep these paths valid.
    \item \textbf{Replanning:} If the current plan is invalid or empty, the Guider invokes the \textbf{Target Selection} algorithm (see \cref{alg:targetselection}) to pick a new $\tau$, and uses the Navigator to compute a new plan $\pi$.
    \item \textbf{Action Selection:} Finally, it restricts the set of available next messages (from the Forecaster) to those that align with the next step in $\pi$.
\end{enumerate}

\section{Case Studies}
\label{sec:casestudy}

\begin{wraptable}[8]{r}{0.4\linewidth}
\vspace{-\baselineskip}
\caption{Descriptive statistics of the evaluated \IO grammars.}
\label{tab:protocol-statistics}
\centering
\small
\rowcolors{2}{gray!20}{white}
\begin{tabular}{@{}lrrrl@{}}
\rowcolor{orange!50}
 & \textbf{Rules} & \textbf{$\Phi$} & $|\mathcal{I}|$ & \textbf{Encoding}\\
\textbf{\SMTP}~\cite{rfc-smtp} & 60 & 2 & 2 & \ASCII\\
\textbf{\FTP}~\cite{rfc-ftp} & 79 & 2 & 4 & \ASCII\\
\textbf{\DNS}~\cite{rfc-dns} & 60 & 7 & 2 & Binary\\
\textbf{\RESTAPI} & 7 & 1 & 2 & \HTTP/\JSON\\
\textbf{\ChatGPT} & 11 & 4 & 2 & \HTTP/\JSON\\

\end{tabular}
\end{wraptable}

To evaluate the versatility of language-based protocol testing, we implemented five distinct protocols using the \IO grammars formalism.
These include three established standard protocols (\SMTP, \FTP, and \DNS) and two modern web-based interactions (\RESTAPI and \ChatGPT).
\Cref{tab:protocol-statistics} provides descriptive statistics for each grammar, including the number of production rules $|R|$, constraints $|\Phi|$, and participating parties $|\mathcal{I}|$.
\Cref{tab:protocol-code} summarizes the auxiliary Python logic required to handle protocol-specific semantics.

\begin{table*}[t]
\caption{Auxiliary Python logic embedded in protocol specifications.}
\label{tab:protocol-code}
\centering
\small
\rowcolors{2}{gray!20}{white}
\begin{tabular}{@{}lrp{10.25cm}@{}}
\rowcolor{orange!50}
\textbf{Protocol} & \textbf{LOC} & \textbf{Semantic Functionality} \\
\textbf{\FTP} & 50 & Dynamic reconfiguration of data channel parties (runtime port switching); state tracking for failed login limits. \\
\textbf{\DNS} & 149 & Message compression/decompression algorithms; domain name encoding; validation of resource records. \\
\textbf{\SMTP} & 14 & \UNIX timestamp conversion; Base64 encoding/decoding for auth. \\
\end{tabular}
\end{table*}

\subsection{SMTP: A Baseline Text Protocol}
\SMTP is the standard protocol for email transmission, defined in \RFC~5321~\cite{rfc-smtp}.
It represents a classic stateful, text-based interaction between two parties: a \textit{Client} and a \textit{Server}.
Our implementation defines a subset of \SMTP focused on unencrypted client authorization and email submission.
The grammar relies exclusively on \ASCII encoding ($\Sigma_{\text{ASCII}}$).
\Cref{fig:smtp-lts} depicts the induced finite-state machine, while \Cref{fig:smtp-grammar} illustrates the structural definitions.

\subsection{FTP: Multi-Party Channels and Dynamic Configuration}
\label{sec:case-study-ftp}

The File Transfer Protocol (\FTP)~\cite{rfc-ftp} introduces significant complexity over \SMTP due to its dual-channel architecture.
\FTP separates \emph{control} information from \emph{data} transfer, effectively requiring the coordination of multiple logical connections.

\begin{wrapfigure}[8]{R}{0.6\linewidth}
\vspace{-0.6\baselineskip}
\renewcommand{\litleft}{‘\bgroup\ulitleft\ttfamily\bfseries}
\begin{densegrammar}\footnotesize
\any <exchange_list> ::= \client<CC:SC:req_list> \server<SC:CC:open_list> \any<list_transfer>

\any <list_transfer> ::= \server<SD:CD:list_data>? \any\big(<finalize_first> | <close_data_first>\big)

\any<finalize_first> ::= \server<SCS:SCC:close_data>\server<SC:CC:finalize_list>

\any<close_data_first> ::= \server<SD:CC:finalize_list>\server<SD:CD:list_data>?<SCS:SCC:close_data>

\end{densegrammar}
\caption{The \FTP \textit{list} command structure.
The command is negotiated over the control channel ($CC \to SC$), while the payload is transmitted via the data channel ($SD \to CD$). $SCS$ and $SCC$ are specification-only parties that model the termination of the data connection.}
\label{fig:ftp-list-grammar}
\end{wrapfigure}

To model this, we define the set of parties as $\mathcal{I} = \{CC, SC, CD, SD\}$, representing the Client and Server endpoints for the Control and Data channels, respectively.
\Cref{fig:ftp-list-grammar} demonstrates how \IO grammars handles channel switching within a single derivation rule: the request is sent via $\pi(\text{\nonterm{req\_list}}) = (CC, SC)$, while the subsequent data transfer is mapped to $\pi(\text{\nonterm{list\_data}}) = (SD, CD)$. 
As \FTP communicates using a Data channel and a Control channel, a message sent using the Control channel may overtake a message sent over the Data channel. A race condition occurs. The helper parties $SCC$ and $SCS$ resolve this race condition by marking the exact point in the protocol execution at which a Data channel is closed, thereby indicating that the data transfer is complete. Neither $SCC$ nor $SCS$ is present in the actual protocol, but just in the specification. They mark the protocol stage where the server is expected to close the data connection.

\begin{wrapfigure}[6]{R}{0.6\linewidth}
\vspace{-1\baselineskip}
\renewcommand{\litleft}{‘\bgroup\ulitleft\ttfamily\bfseries}
\begin{densegrammar}\footnotesize
\any<exchange_epassive> ::= \client<CC:SC:req_epassive> \server<SC:CC:resp_epassive>

\any<resp_epassive> ::= `229 ... (|||' <open_port> `|)\crlf'

\any<open_port> ::= <passive_port> := \textbf{config\_port}(int(<param>))

\any<passive_port> ::= <number> := \textbf{randint}(50000, 50100)
\end{densegrammar}
\caption{Dynamic port negotiation. The \texttt{config\_port} generator binds the grammar value to the underlying socket configuration.}
\label{fig:ftp-port-select}
\end{wrapfigure}

\subsubsection{Dynamic Topology and Semantic Actions}
A critical challenge in \FTP is the \textit{Extended Passive Mode} (EPSV), where the server dynamically selects a TCP port for the data channel and communicates it to the client.
This requires the fuzzer to modify the network configuration of the $CD$ party at runtime based on the parsed input from the $SC$ party.
We address this using \emph{parameterized generators}—semantic actions attached to grammar rules (denoted by \texttt{:=}).
As illustrated in \Cref{fig:ftp-port-select}, the rule \nonterm{open\_port} is bound to the Python function \texttt{config\_port}.
\begin{itemize}
    \item \textbf{When Fuzzing the Client:} The fuzzer acts as the server. It uses \texttt{randint} to select a port, updates its own $SD$ listener, and generates the message containing the port number.
    \item \textbf{When Fuzzing the Server:} The fuzzer acts as the client. Upon parsing the `229` response, the \texttt{config\_port} function is triggered with the parsed integer. This function dynamically reconfigures the $CD$ party to connect to the new port.
\end{itemize}

\subsubsection{Ambiguity and State Superposition}

The \FTP protocol also highlights the challenge of grammatical ambiguity in state transitions.
\Cref{fig:ftp-login-grammar} defines the authentication handshake.
The sequence begins with a username request (\nonterm{req\_login\_user}), which is syntactically identical for both successful and failed login attempts.
Consequently, upon observing this message, the derivation tree $T$ enters a state of \emph{superposition}, simultaneously maintaining valid partial trees for both \nonterm{exchange\_login\_ok} and \nonterm{exchange\_login\_fail}.
The ambiguity is resolved only when the subsequent password response is generated or parsed, collapsing the derivation into the correct branch.

\begin{figure}[t]
\renewcommand{\litleft}{‘\bgroup\ulitleft\ttfamily\bfseries}
\begin{densegrammar}\footnotesize
\any <exchange_login> ::= <exchange_login_ok> | <exchange_login_fail>

\any <exchange_login_ok> ::=
    \client<CC:SC:req_user> \server<SC:CC:resp_user>
    \client<CC:SC:req_pass_ok> \server<SC:CC:resp_pass_ok> \any<state_logged_in>

\any <exchange_login_fail> ::=
    \client<CC:SC:req_user> \server<SC:CC:resp_user>
    \big(
      \client<CC:SC:req_pass_fail> \server<SC:CC:resp_pass_fail> \any<state_retry>
    \big)
\end{densegrammar}
\caption{Subset of the \FTP authentication grammar.
The initial exchange (User/Response) is ambiguous; the path is determined only by the specific password variant (\emph{pass\_ok} vs \emph{pass\_fail}) chosen later.}
\label{fig:ftp-login-grammar}
\end{figure}

\subsection{DNS: Handling Binary Data and Semantic Constraints}
\label{sec:case-study-dns}

\DNS~\cite{rfc-dns} translates human-readable domain names into \IP addresses.
At an abstract level, \DNS is very easy to model---a client sends a domain name to a \DNS server, which returns an \IP address.
The details, however, are much more hairy.
In the \DNS protocol, several header elements and flags are encoded as short bit sequences (\cref{fig:dns-record}) and nontrivial compression schemes.
\FANDANGO supports both generation and parsing of \emph{bit sequences} within strings, which is essential for modeling the \DNS binary format.
Our \IO grammar specification supports \texttt{NS}, \texttt{A}, and \texttt{CNAME} records, showcasing the framework's ability to handle low-level binary data while enforcing complex semantic dependencies.

\newcommand{\zero}{\texttt{\textbf{0}}\xspace}
\newcommand{\one}{\texttt{\textbf{1}}\xspace}

\begin{wrapfigure}[9]{R}{0.45\linewidth}
\vspace{-0.5\baselineskip}
\begin{densegrammar}
<type_a> ::= <type_id_a> <rr_class> <a_ttl> \zero\{13\} \one \zero \zero <ip_address>

<type_id_a> ::= \zero\{15\} \one

<a_ttl> ::= \zero <bit>\{7\} <byte>\{3\}

<ip_address> ::= <byte>\{4\}

<bit> ::= \zero | \one

\end{densegrammar}
\caption[An A record returned by a DNS server]{An \texttt{A} record returned by a \DNS server. The \nonterm{ip\_address} encodes the \IP v4 address returned as a sequence of four bytes.
\zero{} and \one{} represent zero- and one-bits, respectively; the suffix $\{n\}$ denotes $n$ repetitions.}
\label{fig:dns-record}
\end{wrapfigure}

A significant challenge in modeling \DNS is its \textbf{message symmetry}: queries and responses share a nearly identical structural format consisting of a fixed-length header followed by four sections (questions, answers, authority, and additional records). 
This structural overlap limits the ability of a pure context-free grammar to distinguish between syntactically valid and semantically correct interactions based on structure alone. 

To address this, the \DNS~\IO grammar relies on constraints to validate and generate meaningful interactions. 
Specifically, we must ensure that answer records $\server a \any$ align with question records $\client q \any$ based on their \textbf{positional index} within their respective lists. 
We formalize this cross-message dependency in \Cref{eq:dns-constraint}:

\begin{equation}
\label{eq:dns-constraint}
\begin{aligned}
    & \forall \textit{ex} \in \nonterm{start.exchange}, \forall \client q \any \in \textit{ex}.\client\nonterm{dns\_req.question}, \forall \server a \any \in \textit{ex}.\server\nonterm{dns\_resp.answer\_an}\colon \\
    & \quad \text{idx}(\client q\any) = \text{idx}(\server a\any) \implies \\ 
    & \quad \quad \Bigl( \textit{verify\_transitive}(\client q\any, \textit{ex}.\server\nonterm{dns\_resp}\any) \lor 
    (\client q.\nonterm{q\_name}\any = \server a.\nonterm{q\_name\_optional}\any \land \server a.\nonterm{an\_type}[0\!:\!2] = \client q.\nonterm{q\_type}) \Bigr)
\end{aligned}
\end{equation}

In this definition, $\text{idx}(n)$ returns the relative position of a nonterminal within its parent list. 
The constraint ensures that for any question-answer pair sharing the same index, the answer must either be part of a valid transitive chain (e.g., a \texttt{CNAME} resolution) or match the question's name and type directly. 

Because the number of questions and answers can be arbitrary, expressing these positional dependencies is impossible using pure context-free rules, which cannot link nodes across disparate branches of a derivation tree. 
By lifting these requirements into constraints, the \IO grammar successfully models semantic relationships that exist beyond the reach of traditional syntax-directed generation.

\subsection{REST~API: Interfacing with Web Services}
\label{sec:case-study-rest}

Our first custom protocol application evaluates a web-based interaction with a public \RESTAPI. Specifically, the grammar models requests to the \textit{Dune} \API~\footnote{\url{https://github.com/ywalia01/dune-api/tree/v1.0.0}}, which provides movie quotes via \JSON over \HTTP. 
The \IO grammar encapsulates the \HTTP request-response cycle, where the fuzzer acts as the client using the Python \texttt{requests} library.

This study illustrates how \IO grammars can be used to model high-level service interactions where the communication is governed by an \API contract rather than a traditional protocol grammar. 
To ensure logical consistency between the client's request and the server's response, we define a semantic constraint $\phi_{\text{count}}$:
\begin{equation}
\label{eq:rest-constraint}
    \forall \textit{ex} \in \nonterm{start.exchange}\colon \text{val}(\client ex.\nonterm{req.limit}) = \text{count}(\server ex.\nonterm{resp.quotes})
\end{equation}
The constraint ensures that the number of quotes returned in the \JSON payload matches the specific \texttt{limit} parameter requested by the fuzzer. This demonstration highlights the capability of \IO grammars to bridge the gap between structured payloads and semantic state validation in modern web services.

\subsection{ChatGPT: Natural Language Prompt Engineering}
\label{sec:case-study-gpt}

The final case study, presented in \cref{fig:chatgpt-report}, interfaces with \OPENAI's \textit{GPT} models via the official \API. This scenario demonstrates the application of \IO grammars to multi-parameter \API interactions and natural language processing. 
The grammar serves two functions: $(i)$ it selects the model (\texttt{gpt-4.1}), and $(ii)$ it constructs complex prompts that impose specific behavioral constraints on the Large Language Model (LLM).

The grammar generates prompts that instruct the LLM to assume a specific persona and write a report on a subject defined by a stochastic combination of nonterminals: \nonterm{verb}, \nonterm{adjective}, \nonterm{noun}, and \nonterm{place}. 
Furthermore, the prompt includes a \textit{negative constraint}, specifying a character sequence \nonterm{avoid} that must not appear in the output.

\begin{figure}[t]
\centering
\begin{minipage}{0.48\linewidth}
\begin{densegrammar}\footnotesize
\any <exchange> ::= \client<Client:request> \server<Gpt:response>

\any <request> ::= <gpt_model> <prompt>

<gpt_model> ::= `gpt-4.1'

<prompt> ::= `Report about ' <subject> 
             `. Avoid: ' <avoid>

<subject> ::= <verb> <adj> <noun> <place>

<verb> ::= `testing' | `fixing'

<adj> ::= `sustainable' | `innovative'

<noun> ::= `robots' | `rockets'

<place> ::= `on Mars' | `at Google'

<avoid> ::= `crash' | `Elon' | `woke'

<response> ::= \textbf{\ttfamily r}`(?s).*'
\end{densegrammar}
\end{minipage}
\hfill
\begin{minipage}{0.48\linewidth}
\begin{minted}[fontsize=\footnotesize, frame=single]{python}
class Client(Party):
  def send(self, tree, recipient):
    # Extract model and prompt from tree
    model = tree.children[0].to_str()
    prompt = tree.children[1].to_str()
    
    # Call OpenAI API
    res = openai.Chat.create(
            model=model, 
            messages=[{"content": prompt}]
          )
    
    # Pass response back to fuzzer
    self.receive(res.text, sender='GPT')
\end{minted}
\end{minipage}
\caption{The \IO grammar and corresponding Python party implementation for testing \ChatGPT. The implementation extracts generated tokens from the derivation tree to invoke the external \API.}
\label{fig:chatgpt-report}
\end{figure}

The verification of the LLM response is handled by a composite constraint $\Phi_{\text{gpt}}$ that ensures the response (i) contains the subject at some point and (ii) does not contain the forbidden sequence:
\begin{equation}
\label{eq:gpt-constraint}
    \forall \textit{ex} \in \nonterm{start.exchange}\colon 
    (\nonterm{subject} \subset \server ex.\nonterm{response}) \land (\nonterm{avoid} \notin \server ex.\nonterm{response})
\end{equation}
By mapping the \API interaction to a derivation tree, we can use the same evolutionary solver to explore different prompt combinations that might cause the LLM to violate its safety or adherence instructions. 
This case study confirms that \IO grammars are not restricted to established network standards but are equally applicable to self-designed, natural-language-based protocols.

\section{Evaluation}
\label{sec:evaluation}

We evaluate language-based protocol testing by applying it to subjects described in \cref{sec:casestudy}, focusing on the conciseness of the language, the efficiency of our navigation strategy, and the resulting coverage in real-world targets.
We formulate five research questions:

\begin{description}
    \item[RQ1: Conciseness of Specifications.] \emph{How concise are our specifications?}
    We assess the complexity of our specifications in terms of rules and constraints, and the amount of auxiliary Python code required for semantic logic.

    \item[RQ2: Performance.] \emph{What is the performance of language-based protocol testing?}
    This question evaluates the practical footprint of our formalism. We measure the raw throughput of the \FANDANGO engine in producing valid protocol messages.

    \item[RQ3: Coverage of \IIO Grammars.] \emph{To what extent can language-based protocol testing exercise the complete protocol \IO grammar?}
    We evaluate the achieved $k$-path coverage across our subjects. This measures coverage of the \emph{entire interaction}, including fuzzer-controlled messages and observed SUT responses, ensuring that all production rules are exercised at least once.

    \item[RQ4: Impact of Coverage Guidance.] \emph{Does our navigation algorithm achieve grammar coverage significantly faster than unguided generation?}
    We compare the baseline \FANDANGO engine (which explores the grammar through stochastic mutation) against our coverage-guided navigator (\cref{sec:k-path}). We evaluate the \emph{speed of convergence} toward 100\% $k$-path coverage in our test subjects to determine the efficiency of our navigation strategy.

    \item[RQ5: Comparison with Mutation-Based Fuzzing.] \emph{How does \IO grammar-based testing compare to state-of-the-art protocol fuzzers?}
    Using \AFLNET as a baseline for mutation-based protocol fuzzing, we compare the respective code coverage achieved and bugs found.
   
\end{description}

All experiments were conducted on a \texttt{ThinkPad X1 Yoga Gen 8} running \texttt{Ubuntu 24.04.4 LTS}. The system is equipped with a 10-core 13th Gen \texttt{Intel Core i7-1355U} processor (max 5.0 GHz) and 32 GB of LPDDR5 RAM. Fuzzing campaigns were executed within a Docker-based environment to ensure isolation and reproducibility. A guide on how to reproduce the results can be found in the code repository. Our experiments run on a diverse set of target systems across different protocol categories:

\begin{itemize}
    \item \textbf{DNS:} \texttt{BIND 9.20.0-2ubuntu3.1-Ubuntu}.
    \item \textbf{SMTP:} \texttt{OpenSMTPD 7.7.0-portable}.
    \item \textbf{FTP:} \texttt{vsftpd-3.0.5-r2}.
    \item \textbf{REST API:} \textit{Dune API} v1.0.0.
    \item \textbf{LLM:} \textit{GPT-4.1} (via OpenAI API).
\end{itemize}

\subsection{RQ1: Specification Conciseness}
\label{sec:rq1-conciseness}

One of the goals of language-based protocol testing is to reduce the effort required to specify complex interactions. In this research question, we evaluate the \textit{conciseness} of \IO grammars by analyzing their structural complexity and describing the implementation effort relative to other common testing approaches.

\subsubsection{Methodology} 
We perform a structural analysis of the five subject grammars developed for this study.
For each subject grammar, we record the number of production rules ($|R|$), the number of semantic constraints ($|\Phi|$), and the number of participating parties ($|\mathcal{I}|$).

\subsubsection{Conciseness}
Our results are summarized in \Cref{tab:protocol-statistics}.
We observe that our \IO grammar is highly concise, requiring only 12 rules to fully specify the bi-directional interaction (client and server) for the authentication path. This efficiency stems from the grammar's structure: in our models, we generally require \emph{one rule per protocol state} and \emph{one expansion per transition}.
Given that any protocol model must account for these states and transitions, we do not see any way to make the specification more concise than this 1:1 mapping.

\subsubsection{Alternative Realizations}
While a formal comparison of conciseness is difficult to evaluate objectively, our qualitative observations highlight significant differences in effort and capability.
For this comparison, we consider the \FANDANGO \FTP \IO grammar.
Twelve of its rules (= twelve lines of code) are dedicated to the \emph{authentication path,} which we use as a baseline.
\begin{description}
    \item[\PEACH specifications.] A specification in \PEACH~\cite{peach} for the same \FTP authentication features~\cite{proteansec2013fuzzyftp} requires 58~lines for the data and state models alone---versus 12 in the \FANDANGO \IO grammar.
    \PEACH specifications are \emph{asymmetric;} they model the client's perspective to test a server, meaning testing the client would require a separate, inverted model.
    \item[Handwritten code.] We created a ``as short as possible'' handwritten Python fuzzer that covers the \FTP failing authentication path.
    Despite this code
    \begin{enumerate*}[label=(\arabic*)]
        \item again being \emph{asymmetric,} handling the client perspective only;
        \item relying on oversimplified \emph{string-prefix validation} rather than deep structural parsing; and
        \item requiring significant updates should users want to satisfy a particular \emph{constraint,}
    \end{enumerate*}
    it still required 31 lines of code\footnote{\url{https://gist.github.com/alex9849/2a71fecd3b375aa854cac44ee2164f26}}---again compared against the 12~lines in the \FANDANGO \IO grammar.
    \item[Parser and producer specifications.] Being a grammar, the \FANDANGO \IO grammar can also serve as a concise specification for parser generators like \ANTLR{}~\cite{antlr} or producers like \GRAMMARINATOR~\cite{hodovan2018grammarinator} and then have roughly the same length.
    But then, the whole \emph{interaction} aspect would be gone: a standalone parser or producer could not react to the messages being received.
    Semantic constraints, if at all, would have to be hard-coded into the parser or producer code.
\end{description}

Crucially, the additional \textit{constraints} ($\Phi$) in an \IO grammar allow for validation of semantic properties that a standard context-free parser cannot capture.
By centralizing syntax, semantics, and interaction in a single grammar, we eliminate the need for dozens to thousands of lines of auxiliary \emph{glue code} for managing network I/O and synchronizing state.

\conclusion{Language-based protocol testing uses minimal, unified specifications that~map~directly~to~protocol~states and~transitions.}

\subsection{RQ2: Performance}
\label{sec:rq2-performance}

\begin{wraptable}[8]{r}{0.5\linewidth}
\vspace{-\baselineskip}
\caption{Performance metrics for \IO grammar subjects over 1-hour runs. \textit{T-100\%} denotes the time in seconds to achieve full k-path coverage.}
\label{tab:performance-results}
\centering
\small
\rowcolors{2}{gray!20}{white}
\begin{tabular}{@{}lrrrrr@{}}
\rowcolor{orange!50}
\textbf{Subject} & \textbf{T-100\%} & \textbf{Out} & \textbf{In} & \textbf{Total} & \textbf{Msg/s} \\
\textbf{\SMTP} & 19.94 & 16,516 & 17,361 & 33,877 & 9.4 \\
\textbf{\FTP} & 416.33 & 8,375 & 9,477 & 17,852 & 5.0 \\
\textbf{\RESTAPI} & 4.18 & 6,387 & 6,387 & 12,774 & 3.5 \\
\textbf{\DNS} & -- & 546 & 543 & 1,089 & 0.3 \\
\end{tabular}
\end{wraptable}

\textbf{What is the performance of language-based protocol testing?}
To evaluate the performance of language-based protocol testing, we measure the \emph{message throughput} across our case studies. This experiment quantifies the engine's ability to sustain efficient fuzzing while adhering to complex \IO grammar structures and semantic constraints.

\subsubsection{Methodology} 
For each of the subjects (\SMTP, \FTP, \DNS, and \textit{\RESTAPI}), we exercise the protocol for a duration of one hour. The execution is split into two distinct phases: 
\begin{enumerate} 
    \item \textbf{Exploration Phase:} The fuzzer employs our coverage-guided navigation strategy (\cref{sec:navigation}) until 100\% $k$-path grammar coverage ($k=5$) is achieved. We record the time required to reach this saturation point. 
    \item \textbf{Exploitation Phase:} Once 100\% coverage is reached, the fuzzer switches to an unguided, stochastic generation mode for the remainder of the hour. This phase maximizes the volume of inputs to explore the broader input space within the established protocol states. 
\end{enumerate}

We distinguish between \emph{outgoing} messages (generated by the fuzzer) and \emph{incoming} messages (received and parsed from the SUT). Notably, incoming messages typically involve higher latency as they require network transmission and processing by the external system.
We excluded \ChatGPT from this specific performance benchmark due to the costs associated with high-volume API usage.

\subsubsection{Analysis of Results} 
As shown in \Cref{tab:performance-results}, the throughput varies significantly depending on the protocol's transport layer and state complexity.

\begin{itemize}
    \item \textbf{High Throughput (\SMTP, \FTP):} The stateful text-based protocols running on local {\smaller TCP} connections achieved the highest throughput, with \SMTP reaching approximately 9.4 messages per second. This demonstrates that the overhead of the \IO grammar engine (producing, parsing and state tracking) remains low even under sustained protocol activity.
    
    \item \textbf{Latency-Bound (\REST):} The \textit{Dune} API, being a stateless {\smaller HTTP}-based interaction, achieved a consistent throughput of 3.5 msg/s. Here, the bottleneck is primarily the {\small HTTP} round-trip time rather than the generation logic.
    
    \item \textbf{Timeouts and Errors (\DNS):} \DNS exhibited the lowest throughput (0.3 msg/s). During the experiments, we encountered both \texttt{SERVFAIL} responses and request timeouts (e.g., \texttt{dig CNAME hansen.info}), which introduced additional waiting periods. The primary performance cost, however, stems from the \emph{early parser}, which relies on flexible bitwise parsing of binary \DNS responses. While this design provides high adaptability across diverse inputs, it incurs higher parsing overhead. This limitation is inherited from the original \FANDANGO framework rather than from language-based protocol testing itself. Future work could improve performance by generating language-tailored parsers (e.g., with \ANTLR) during grammar preprocessing. Additionally, \DNS only reached 83.08\% coverage, as the server implementation did not support the full range of response types defined in our general grammar. Full input grammar coverage was achieved within 5.6 seconds.
\end{itemize}

\conclusion{Language-based protocol testing sustains viable throughput for fuzzing campaigns. Performance is largely bound by the network latency and SUT response times.}

\subsection{RQ3: Interaction Grammar Coverage}
\label{sec:rq3-coverage}

\textbf{To what extent can language-based protocol testing exercise the complete protocol \IO grammar?} We evaluate the capacity of our approach to explore and satisfy the protocol \IO grammar. 
Unlike standard grammar-based fuzzing, which only generates, \IO grammars require a synchronized interaction between the fuzzer and the system under test (SUT) to reach deep structural states.

\subsubsection{Methodology} 
We evaluate our subjects (\DNS, \SMTP, \FTP, \ChatGPT, and the \textit{Dune} \API) using $k$-path coverage as the primary metric. 
We configure the fuzzer to target a path length of $k=5$, which implies exploring combinations of alternatives, states, and transitions up to five levels deep. 
We perform ten independent runs for each subject in two scenarios:

\begin{enumerate}
    \item \textbf{Guided:} Using our coverage-driven navigation strategy (\cref{sec:navigation}).
    \item \textbf{Random:} Selecting admissible message types randomly without coverage feedback.
\end{enumerate}
Progress is tracked over a logarithmic timescale to capture the initial expansion and the saturation of the grammar space.

\begin{figure}[ht]
  \centering
  \hspace{0.74cm}
\begin{tikzpicture}
  \begin{axis}[
    width=\textwidth-0.68cm,
    height=2cm,
    ticks=none,
    xtick=\empty,
    ytick=\empty,
    axis line style={draw=none},
    legend columns=7,
    legend style={
      draw=none,
      fill=none,
      at={(1.0,0.5)},
      anchor=east,
      font=\footnotesize,
      align=center
    }
  ]
      \addplot[draw=none, forget plot] coordinates {(0,0)};

    \addlegendimage{draw=black, ultra thick, mark=none}
    \addlegendentry{Guided}
    \addlegendimage{draw=black, ultra thick, dotted, mark=none}
    \addlegendentry{Random $\qquad\qquad\qquad$}
    \addlegendimage{draw=blue, ultra thick, mark=none}
    \addlegendentry{\SMTP}
    \addlegendimage{draw=red, ultra thick, mark=none}
    \addlegendentry{\FTP}
    \addlegendimage{draw=ForestGreen, ultra thick, mark=none}
    \addlegendentry{\DNS}
    \addlegendimage{draw=Orange, ultra thick, mark=none}
    \addlegendentry{\ChatGPT}
    \addlegendimage{draw=Purple, ultra thick, mark=none}
    \addlegendentry{Dune}
  \end{axis}
\end{tikzpicture}
  \vspace{0.5em}
 
  \begin{subfigure}{0.49\textwidth}
    \centering
    \begin{tikzpicture}
      \begin{axis}[
        enlarge x limits=false,
        grid=major,
        grid style={dashed,black!30},
        legend style={at={(0.993,0.02)},anchor=south east},
        legend columns=3,
        width=\textwidth,
        height=6cm,
        xlabel={Time [Seconds]},
        ylabel={$k$-path Coverage [\%]},
        xmode=log,
        log basis x=10,
        xmin=1,
        xmax=10000,
        xtick={1,10,100,1000,10000},
        xticklabels={1,10,100,1000,10000},
        log ticks with fixed point
      ]
        \addplot[very thick, blue, forget plot]
          table[x=time, y expr=\thisrow{meanpercentroleuniqueClient}*100,col sep=comma]
          {tikz/msgs/smtp\string_median\string_grammar\string_coverage\string_guided.csv};

        \addplot[very thick, dotted, blue, forget plot]
          table[x=time, y expr=\thisrow{meanpercentroleuniqueClient}*100,col sep=comma]
          {tikz/msgs/smtp\string_median\string_grammar\string_coverage\string_unguided.csv};

        \addplot[very thick, red, forget plot]
          table[x=time, y expr=\thisrow{meanpercentroleuniqueClientControl}*100,col sep=comma]
          {tikz/msgs/ftp\string_median\string_grammar\string_coverage\string_guided.csv};
        \addplot[very thick, dotted, red, forget plot]
          table[x=time, y expr=\thisrow{meanpercentroleuniqueClientControl}*100,col sep=comma]
          {tikz/msgs/ftp\string_median\string_grammar\string_coverage\string_unguided.csv};

        \addplot[very thick, ForestGreen, forget plot]
          table[x=time, y expr=\thisrow{meanpercentroleuniqueClient}*100,col sep=comma]
          {tikz/msgs/dns\string_median\string_grammar\string_coverage\string_guided.csv};
        \addplot[very thick, dotted, ForestGreen, forget plot]
          table[x=time, y expr=\thisrow{meanpercentroleuniqueClient}*100,col sep=comma]
          {tikz/msgs/dns\string_median\string_grammar\string_coverage\string_unguided.csv};

        \addplot[very thick, Orange, forget plot]
          table[x=time, y expr=\thisrow{meanpercentroleuniqueClient}*100,col sep=comma]
          {tikz/msgs/chatgpt\string_median\string_grammar\string_coverage\string_guided.csv};
        \addplot[very thick, dotted, Orange, forget plot]
          table[x=time, y expr=\thisrow{meanpercentroleuniqueClient}*100,col sep=comma]
          {tikz/msgs/chatgpt\string_median\string_grammar\string_coverage\string_unguided.csv};

        \addplot[very thick, Purple, forget plot]
          table[x=time, y expr=\thisrow{meanpercentroleuniqueClient}*100,col sep=comma]
          {tikz/msgs/dune\string_median\string_grammar\string_coverage\string_guided.csv};
        \addplot[very thick, dotted, Purple, forget plot]
          table[x=time, y expr=\thisrow{meanpercentroleuniqueClient}*100,col sep=comma]
          {tikz/msgs/dune\string_median\string_grammar\string_coverage\string_unguided.csv};
      \end{axis}
    \end{tikzpicture}
    \caption{\emph{Message} coverage over time}
    \label{fig:client-msg-coverage-time}
  \end{subfigure}
  \hfill
    \begin{subfigure}{0.49\textwidth}
    \centering
    \begin{tikzpicture}
      \begin{axis}[
        enlarge x limits=false,
        grid=major,
        grid style={dashed,black!30},
        legend style={at={(0.993,0.02)},anchor=south east},
        legend columns=3,
        width=\textwidth,
        height=6cm,
        xlabel={Time [Seconds]},
        ylabel={},
        xmode=log,
        log basis x=10,
        xmin=1,
        xmax=10000,
        xtick={1,10,100,1000,10000},
        xticklabels={1,10,100,1000,10000},
        log ticks with fixed point
      ]
        \addplot[very thick, blue, forget plot]
          table[x=time, y expr=\thisrow{meanpercentstart}*100,col sep=comma]
          {tikz/all/smtp\string_median\string_grammar\string_coverage\string_guided.csv};

        \addplot[very thick, dotted, blue, forget plot]
          table[x=time, y expr=\thisrow{meanpercentstart}*100,col sep=comma]
          {tikz/all/smtp\string_median\string_grammar\string_coverage\string_unguided.csv};

        \addplot[very thick, red, forget plot]
          table[x=time, y expr=\thisrow{meanpercentstart}*100,col sep=comma]
          {tikz/all/ftp\string_median\string_grammar\string_coverage\string_guided.csv};
        \addplot[very thick, dotted, red, forget plot]
          table[x=time, y expr=\thisrow{meanpercentstart}*100,col sep=comma]
          {tikz/all/ftp\string_median\string_grammar\string_coverage\string_unguided.csv};

        \addplot[very thick, ForestGreen, forget plot]
          table[x=time, y expr=\thisrow{meanpercentstart}*100,col sep=comma]
          {tikz/all/dns\string_median\string_grammar\string_coverage\string_guided.csv};
        \addplot[very thick, dotted, ForestGreen, forget plot]
          table[x=time, y expr=\thisrow{meanpercentstart}*100,col sep=comma]
          {tikz/all/dns\string_median\string_grammar\string_coverage\string_unguided.csv};

        \addplot[very thick, Orange, forget plot]
          table[x=time, y expr=\thisrow{meanpercentstart}*100,col sep=comma]
          {tikz/all/chatgpt\string_median\string_grammar\string_coverage\string_guided.csv};
        \addplot[very thick, dotted, Orange, forget plot]
          table[x=time, y expr=\thisrow{meanpercentstart}*100,col sep=comma]
          {tikz/all/chatgpt\string_median\string_grammar\string_coverage\string_unguided.csv};

        \addplot[very thick, Purple, forget plot]
          table[x=time, y expr=\thisrow{meanpercentstart}*100,col sep=comma]
          {tikz/all/dune\string_median\string_grammar\string_coverage\string_guided.csv};
        \addplot[very thick, dotted, Purple, forget plot]
          table[x=time, y expr=\thisrow{meanpercentstart}*100,col sep=comma]
          {tikz/all/dune\string_median\string_grammar\string_coverage\string_unguided.csv};
      \end{axis}
    \end{tikzpicture}
    \caption{\emph{Grammar} coverage (including responses) over time}
    \label{fig:grammar-coverage-time}
  \end{subfigure}
  \vspace{-0.5\baselineskip}
  \caption{$k$-path coverage over time with $k=5$; note the logarithmic x-axis $\log_{10}(t)$. Solid lines represent $k$-path guidance; dotted lines represent unguided (random) testing.}
  \label{fig:two-graphs}
\end{figure}

\subsubsection{Input Space Exploration} 
We first evaluate the \emph{message coverage}---the set of nonterminals and production rules associated with fuzzer-controlled parties ($\mathcal{I}_{\text{fuzz}}$). 
It is important to note that achieving 100\% coverage is not guaranteed by construction. 
In \IO grammars, certain message branches are guarded by specific SUT responses; if the fuzzer fails to trigger those responses or if semantic constraints ($\Phi$) are not solved, those branches remain unreachable. 

As shown in \Cref{fig:client-msg-coverage-time}, both guided and random variants achieve 100\% coverage across all subjects. 
This rapid saturation confirms that the engine effectively explores the input alternatives defined in the grammar.

\subsubsection{Full Interaction Space Exploration} 
The core strength of \IO grammars lies in exercising the \emph{entire} grammar, including the responses received from the SUT ($\mathcal{I}_{\text{ext}}$). \Cref{fig:grammar-coverage-time} illustrates the total grammar coverage over time. 

For most subjects, 100\% total interaction coverage was achieved, indicating that the subjects covered all specified interactions and the fuzzer successfully generated all prerequisite traffic. 
The notable exception is \DNS, which reached~81.92\% interaction coverage. This shortfall is not a failure of the fuzzer but a reflection of the \texttt{bind9} implementation, which does not produce certain response variants defined in our generalized \IO grammar. This illustrates that interaction coverage is always relative to the intersection of the grammar and the SUT's actual behavior.

\conclusion{Language-based protocol testing enables exhaustive exploration of the entire defined interaction space, including all specified messages, responses, and state transitions.}

\subsection{RQ4: Impact of Coverage Guidance}
\label{sec:rq4-guidance}

\textbf{Does our navigation algorithm achieve grammar coverage significantly faster than unguided generation?} We evaluate the efficiency of our coverage-guided navigation strategy and investigate whether our algorithm can accelerate the discovery of deep protocol states.

\subsubsection{Methodology}
We conduct an ablation study by comparing the coverage-guided navigator (\cref{sec:navigation}) against a baseline unguided generator. The unguided version selects among admissible production rules with uniform probability, essentially performing a random walk through the interaction space. We use the same subjects and the same path length $k=5$ as in \cref{sec:rq3-coverage}, measuring the time required to reach saturation for both methods.

\subsubsection{Convergence Speed}

As we see in \Cref{fig:client-msg-coverage-time,fig:grammar-coverage-time}, the coverage-guided navigator reaches the same coverage substantially faster than the baseline across all subjects.
For protocols with simple state models, such as the \textit{Dune} \API or \ChatGPT, both approaches converge quickly. However, as the complexity of the interaction increases, the benefit of purposeful navigation becomes more pronounced.

On average, our guided approach achieves full message coverage 52.12\% faster than the unguided approach.
On average, our guided approach achieves full \IO coverage 61.21\% faster than the unguided approach.
This is because our navigation algorithm actively prioritizes paths leading to unexplored nonterminals, whereas the random walker often becomes trapped in ``shallow'' loops (e.g., repeating simple command-response cycles).

\subsubsection{Deep State Reachability}
The most significant result is observed in the \FTP subject. The coverage-guided variant achieved 100\% interaction coverage in approximately 8~minutes. In contrast, the unguided variant achieved 100\% coverage after 3.0~hours of continuous execution. 
Specifically, random testing failed to reach the deeply nested state required to execute the \texttt{LIST} command in passive mode using a binary transfer format. 

This confirms that for complex, stateful protocols, systematic exploration is not merely an optimization but a \emph{requirement} for completeness.
Without guidance, the probability of choosing the specific sequence of actions and parameters needed to trigger rare protocol behaviors becomes vanishingly small.

\conclusion{Navigation guided by \IO grammar coverage reaches deep protocol states significantly faster than random testing.}

\subsection{RQ5: Comparison with Mutation-Based Fuzzing}
\label{sec:rq5-aflnet}

\textbf{How does language-based protocol testing compare to state-of-the-art protocol fuzzers?}
We investigate how language-based protocol testing compares to and complements state-of-the-art mutation-based protocol fuzzers. Specifically, we evaluate whether our approach can overcome the structural fragility of random mutations in the presence of complex protocol languages.

\subsubsection{Methodology}

We select \AFLNET~\cite{pham2020aflnet}, a widely used and continuously maintained coverage-guided greybox fuzzer for network protocols, as our baseline.
Our experiment focuses on the \texttt{lightftp} \FTP server (commit \texttt{5980ea1}), the benchmark used by the \AFLNET authors~\cite{pham2020aflnet}.
We compare two scenarios:
\begin{description}
    \item[\AFLNET Baseline:] We execute \AFLNET using its original, default seed corpus provided by the authors for the \FTP protocol.
    To ensure \AFLNET can log in to the \FTP server, we added one more seed to the corpus containing admin credentials for the \FTP server.
    \item[\IIO Grammar:] The \AFLNET dictionary contains \emph{all \FTP commands available} in \texttt{lightftp}; hence, \AFLNET can cover all commands.
    For a fair comparison, we extended the \FTP \IO grammar with the same complete command input space of the \FTP server.
\end{description}
In both scenarios, we modified the configuration of \texttt{lightftp} to allow both tools to optionally run their connections via \TLS (\FTPS).

We measured statement and branch coverage in \texttt{lightftp} using \texttt{gcov}.
We ran \AFLNET 10~times for 24 hours; \FANDANGO was run until it reached 100\% grammar coverage, which took $\sim$13.5 minutes on average.

\subsubsection{Comparing Coverage}

Our results show that \AFLNET achieved a median statement coverage of 69.6\% and a median branch coverage of 49.7\%.
In contrast, \FANDANGO using \IO grammars accomplishes 80.3\% statement coverage and 52.4\% branch coverage.
Note again that \FANDANGO completes its job when reaching 100\% grammar coverage, after 13.5 minutes.

\conclusion{Language-based protocol testing completes achieves a 80.3\% statement coverage, \\ higher than the 69.6\% statement coverage of \AFLNET, and does so in less than 1\% of the time.}

\subsubsection{Differences in Coverage}

The \AFLNET and \FANDANGO tools cover different code, highlighting the differences between both approaches:
\begin{description}
    \item[Code covered by \AFLNET only.] In contrast to \FANDANGO with \IO grammars, \AFLNET explored more \emph{error-handling paths,} such as permission failures or invalid arguments.
    This is expected, as \FANDANGO produces valid inputs and interactions by construction, and thus exercises fewer to no error-handling paths.
    \item[Code covered by \FANDANGO only.] \FANDANGO exclusively exercised the following code:
    \begin{enumerate}
    \item Code handling traffic on the \emph{FTP data channel.}
    \AFLNET could not reach this code because it targets only one channel per fuzzing run.
    Also, \AFLNET can not draw on any logic to dynamically obtain the data port from the \FTP server and connect to it.
    \item Code handling \emph{handshakes on existing sockets} for \TLS-encrypted messages, for which \AFLNET did not generate correct commands.
    \end{enumerate}
   
\end{description}
In principle, one could use the \FANDANGO-generated interactions as \emph{seeds} for \AFLNET, thus covering the best of both worlds.
However, it is difficult to imagine a \emph{generic} fuzzing approach that would be able to infer the complex and dynamic interactions of \FTP without any form of specification.\footnote{Note that \AFLNET also comes with a \emph{dictionary} of valid \FTP commands, which is a specification already.}

\conclusion{\AFLNET triggers several error-handling paths, \\ while \FANDANGO covers complex and dynamic interactions of the \FTP protocol.}

\subsubsection{Bugs Found}

During fuzzing \texttt{lightftp} with \FANDANGO, we identified a previously unknown bug related to switching the protection level of the data channel during an existing TLS session:
The \FTP \texttt{PROT C} command is supposed to switch communication on the data channel from encrypted to unencrypted.
\texttt{lightftp} acknowledges the command with a 200 return code, yet sticks to encrypted communication.\footnote{We reported this as \texttt{lightftp} issue \#67.}

Mutation-based fuzzing approaches do not support program- or protocol-specific oracles, but aim for generic failure like a crash or a hang.
In contrast, this bug could only be discovered because the \IO grammar encodes the expected (correct) behavior.
By construction, the ability to find \emph{missing} or \emph{incorrect} behavior during testing requires a specification to compare against, and language-based protocol testing provides exactly that.

\conclusion{Language-based protocol testing can detect missing and incorrect behavior with~respect~to~the~specification.}

\subsection{Threats to Validity}
\label{sec:threats}

While our evaluation demonstrates the effectiveness of language-based protocol testing across multiple dimensions, we acknowledge the threats to the validity of our results. 
We categorize these threats into external, construct, and internal validity.

\begin{description}
    \item[External validity.] 
    Our evaluation is based on a set of five subject protocols. 
    While we selected these subjects to cover a diverse range of categories (including binary protocols, text-based protocols, and natural language interfaces), they may not represent the full complexity of all possible network interactions.

    Our comparison with existing mutation-based fuzzers (\Cref{sec:rq5-aflnet}) is limited to one fuzzer (\AFLNET) and one subject only.
    While other tools and subjects may obtain different results, we see a clear causality between the unique features of our approach and the effects we observe.

    \item[Construct validity.]
    A potential threat lies in our choice of evaluation metrics.
    We primarily relied on \emph{$k$-path grammar coverage} (\cref{sec:rq3-coverage}) rather than code coverage to measure the effectiveness of \IO grammars.
    In a specification-based setting, code coverage is a confounding variable: it is largely determined by the \emph{completeness} of the user-written grammar rather than the performance of the test generator.
    For instance, our \DNS grammar intentionally omits \texttt{TXT} records; thus, the corresponding code in \texttt{bind9} is unreachable regardless of the fuzzer's efficiency.
    To mitigate this, we focused on evaluating how thoroughly our method explores the \emph{defined interaction space}, leveraging the established correlation that higher $k$-path coverage induces higher code coverage within the specified scope.

    Comparing code coverage (\Cref{sec:rq5-aflnet}) assumes that all code is similarly likely to have bugs and similarly valuable to cover, which is a simplifying assumption.
    Error-handling paths, for instance, may be less or more valuable than actual functionality, depending on the perspective.

    \item[Internal validity.] 
    Comparing \FANDANGO against existing tools presents challenges due to fundamental differences.
    Any direct comparison against mutation-based fuzzers is problematic, because their performance depends on a large number of factors:
    mutation-based fuzzers are determined by the quality of initial seeds and the used dictionary, whereas \FANDANGO depends on the quality of the \IO grammar;
    mutation-based fuzzers are guided by \emph{code coverage,} whereas \FANDANGO is guided by \emph{specification coverage}.
    In our evaluation (\Cref{sec:rq5-aflnet}), we partially mitigated this by providing \AFLNET with extended seeds; however, a comparison controlling all possible factors is beyond the scope of this paper.
\end{description}

\section{Related Work}
\label{sec:related-work}

\subsection{Specifying Protocols}

Protocols are typically specified using a mix of traditional techniques.
\emph{Finite-state machines} such as \emph{labeled transition systems} are often used to describe the behavior of individual parties, modeling the states of a party and the transitions between them based on received messages.

\begin{description}
  \item[Formal verification.] As \emph{state models} abstract away details of individual messages, state machines also often form the base for formal verification techniques.
\emph{Model checkers} like \SPIN~\cite{spin} or \NuSMV~\cite{nusmv} can be used to verify the correctness of a protocol by checking if the model satisfies certain properties, but are typically limited to a single party.
\TAMARIN~\cite{tamarin} verifies security properties of protocols (including multiple parties), using a dedicated language for cryptographic properties.
  \item[Model-driven testing.] \emph{Model-driven testing} derives abstract test suites from models (typically state machines), which then map into concrete test cases~\cite{neto2007mbt}.
Microsoft's \emph{Spec Explorer}~\cite{veanes2008specexplorer} provides a domain-specific language to specify abstract state machines, including \emph{model programs} that define the conditions under which a transition can be taken.
When instantiated into test cases for reactive systems, parameters of these model programs are instantiated according to their type or using user-chosen values.
Tretmans et al.~\cite{tretmans1992-conformance-testing} generate test suites from \emph{labeled transition systems.}
The approach synthesizes test cases in \TTCN notation.
These test cases execute specification traces and then check whether the implementation's outputs are permitted by the specification, but are limited to the capabilities of \TTCN.
\item[Session types~\textnormal{\cite{castagna2009sessiontypes, scalas2019sessiontypes}}] define the structure and correctness of message exchanges between communicating parties, but lack syntax or complex semantic properties of messages being exchanged; and thus are too limited for our testing purposes.
\end{description}

In contrast, \IO grammars not only describes the states of all parties in a protocol (typically, by embedding them into the grammar, as demonstrated in \Cref{fig:smtp-lts}), but also the \emph{syntax of the individual messages} exchanged between them; the \emph{semantic constraints} capture encodings, conditions, and other semantic properties.
Hence, we provide not just a \emph{model} of a protocol, but a \emph{full executable specification} encompassing all layers of the protocol stack.

\subsection{Specifying Languages}

Language specifications describe the syntax of messages exchanged between parties.

\begin{description}
  \item[Grammars] are a well-known mechanism for specifying the syntax of human-readable messages.
  \RFC documents, for instance, frequently use grammars in augmented Backus-Naur Form~\cite{rfc5234}.
  \item[C-like data structures] are often used to specify the syntax of binary messages.
  \XDR~\cite{rfc4506} specifies binary data for \RFC documents.
  \emph{Binary templates} for the 010 binary editor~\cite{010editor} combine C-like data structures with additional parsing and decomposition functions.
\end{description}

All these specifications cater to individual files and messages and do not capture protocol interactions.
However, they are equivalent to context-free grammars, and thus can be easily embedded in \IO grammars; our \emph{constraints} provide a way to formally express semantic features.

\subsection{Parsing and Producing}

\iffalse 
To \emph{parse} messages, one can either write a parser from scratch, or use a parser generator such as \ANTLR~\cite{antlr}, which takes a grammar as input.
\emph{Producing} messages from grammars is generally straight-forward; the \GRAMMARINATOR~\cite{hodovan2018grammarinator} tool produces inputs from feature-rich \ANTLR specs.
None of these tools can handle semantic features or protocol interactions; they are also limited to either parsing or producing.
\fi

A number of tools combine aspects of \emph{parsing} and \emph{producing} messages for testing and fuzzing.

\begin{description}
  \item[\PEACH.] Close to our approach is the \PEACH protocol fuzzer~\cite{peach}.
\PEACH specifications, known as \emph{\PEACH pits,} define a \emph{data model} specifying the syntax of protocol messages through field types, lengths, and data relationships; a \emph{state model} describes the states of a protocol and transitions between them.
In contrast to \IO grammars, a \PEACH pit can only model a single party.
Hence, clients and servers must be specified independently, and their interaction is not captured in a single specification.
Semantic features are limited to built-in \emph{fixups} that can be used to compute field values from other fields.
Finally, \PEACH offers no support for systematic coverage of interactions; these have to be implemented manually~\cite{peach}.
  \item[Protocol fuzzing libraries.] Fuzzowski~\cite{fuzzowski} is a Python library for fuzzing network protocols, extending the capabilities of BooFuzz~\cite{boofuzz} and Sulley~\cite{sulley}.
It sends \emph{Request} objects to a party and can parse \emph{responses} for certain elements such as authorization tokens.
\emph{Monitors} can test if a fuzzing target still reacts to inputs.
Such libraries greatly ease the task of programming a fuzzer.
However, they lack all declarative aspects of an independent protocol specification, as in \IO grammars and semantic constraints.
  \item[Protocol-specific fuzzers.] Besides these general-purpose tools, a large number of protocol-specific fuzzing tools exist~\cite{zhang2024survey}.
They only test a single party in isolation (albeit well so, being specialized tools), and are not based on a formal specification of the protocol.
\end{description}

\subsection{Fuzzing Messages}

Rather than generating interactions from scratch, one can also \emph{mutate} existing interactions.

\begin{description}
  \item[AutoFuzz~\textnormal{\cite{gorbunov2010autofuzz}}] is a \emph{man-in-the-middle fuzzer:} Given a client and a server, AutoFuzz applies fuzz testing to protocols by intercepting and mutating messages.
 
  \item[\AFLNET{}~\textnormal{\cite{pham2020aflnet}}] is a stateful fuzzer that operates on the client side and relies on server response codes to detect failures and guide fuzzing.
It starts from a seed corpus of request-response traces and uses \AFL to mutate requests guided by code coverage.
  \item[\StateAFL{}~\textnormal{\cite{stateAfl}}] extends \AFLNET by also analyzing long-lived memory to infer and cover states in the system under test.
  It was last updated in 2022, and no longer appears to be actively maintained.
  \item[\FuzztructionNET~\textnormal{\cite{fuzztruction-net}}] mutates \emph{output} code of existing protocol parties to mutate interactions, thus increasing the chances of syntactic and semantic correctness.
\end{description}

As true fuzzers, these are great for testing the robustness of parsers and discovering generic bugs such as crashes or hangs.
But again, in contrast to \IO grammars, they are limited to testing one party (typically the server).
As we show in \Cref{sec:rq5-aflnet}, such fuzzers will have difficulties discovering relationships between messages, such as an \FTP server communicating a port to be used for the data channel.
Furthermore, an initial population of real-world messages may induce a \emph{bias} in testing that may skip rare or edge-case protocol features that \IO grammars cover by construction.
The biggest difference, though, is that \emph{we go beyond simple fuzzing:} We aim at \emph{testing} implementations against adherence to the full protocol, and thus allow \emph{oracles} as constraints that identify and check individual message elements.
Such checking is only possible with a complete formal specification of the protocol, which \IO grammars and semantic constraints provide.

\subsection{Interaction Grammars}

The reasons we choose \emph{grammars} as a foundation for specifying and testing protocols are straight-forward:
\begin{enumerate}
  \item Grammars are a well-studied mechanism for specifying the \emph{syntax of messages};
  \item Grammars subsume regular languages, and hence their representation as \emph{state models}; and
  \item Grammars, like all formal language specifications, can be used for both \emph{parsing} and \emph{producing} strings.
  \item In our \emph{language-based} context, grammars can be combined with constraints over grammar elements to express complex semantic features of protocols, such as encodings, conditions, or relationships between messages and states.
\end{enumerate}
In principle, all these should make grammars a great formal foundation for any formalism that attempts to specify and test protocols.
Yet, such a unified approach is hardly ever found in the literature.

Close to our approach of \emph{\IO grammars} are Real-Time Asynchronous Grammars (\RTAG{}s)~\cite{anderson88-rtag},
attribute grammars that specify stateful protocols.
They support a number of features not found in \IO grammars, such as timers and concurrent productions.
The \RTAG system compiles them into executable protocol \emph{implementations} with efficient parsers, thus achieving correctness by construction.
\RTAG, however, was designed as a formalism for \emph{creating,} not \emph{testing} implementations:
The actual \emph{production} of message elements requires hand-coded interface routines, and there is no concept of systematic coverage.

In the context of software testing, Jones, Harman, and Danicic~\cite{jones1999iogrammars}
observed that a grammar to produce an input for a (stateless) program under test could be combined with a second grammar to parse its output;
a so-called \emph{input/output grammar} would then combine the two into one.
Our \IO grammars also feature this duality of production and parsing, but go much further than just input and output; they capture much more complex interactions between multiple parties, including embedded state models.

The term ``interaction grammar'' is not to be confused with a concept of the same name in \emph{computational linguistics.}
Here, an ``interaction grammar'' is a grammatical formalism for natural languages like English and French, modeling the distinction between saturated and unsaturated syntactic structures~\cite{guillaume2008interactiongrammars}.

\iffalse
    \item[RTAG.~\cite{anderson88-rtag}]  is a formalism for specifying stateful protocols based on attributed grammars. These grammars can be translated into executable protocol implementations. \RTAG was not designed as a testing formalism. It treats grammars as descriptive control structures to derive implementations of the protocol. In contrast to them, our approach treats grammars as executable models for driving and guiding protocol testing. While our approach also derives a protocol implementation from the specification by interpreting it, does our approach also lead the protocol session into states that are underexplored according to our grammar coverage metric. Our approach, which inherits the grammars from Fandango, allows more complex message semantics. In RTAG, which is based on attributed grammars, symbol semantics can only depend on tree nodes on the left-hand side of a production rule, constraining attribute evaluation to a left-to-right derivation order and forbidding dependencies on later symbols.
\fi

\section{Conclusion and Future Work}
\label{sec:conclusion}

Protocol testing is hard, in particular in the absence of a formal specification.
We introduce \emph{language-based protocol testing}, the first approach to specify, automatically test, and systematically cover the full state and input space of protocol implementations.
Language-based protocol testing is based on three ingredients:
\begin{enumerate}%[label=(\arabic*)]
  \item A novel grammar flavor, called \emph{\IO grammar,} specifies the behavior of all parties in a protocol and hence can be used to test either clients or servers.
  \item Using \emph{semantic constraints} from language-based testing, we can express arbitrary predicates and relationships over grammar elements (including messages and states).
  \item Finally, the \emph{\FANDANGO infrastructure} gives us the means to systematically cover all grammar elements (again involving messages and states) and to efficiently solve the specified constraints---features so far not found in other protocol testing tools.
\end{enumerate}
We demonstrate the usefulness of this combination by providing five highly detailed specifications of real-world protocols, including \FTP, \SMTP, and \DNS, and show how to use them for testing real-world implementations of clients and servers.
We show that a single \IO grammar is sufficient to model the full behavior of all involved protocol parties, and consequently, to test the other parties; indeed, our specifications conform to real-world implementations (and vice versa).
By systematically covering input and state space, we obtain a much higher code coverage in programs under test.

Our future work will focus on the following aspects:
\begin{description}
   
    \item[Coverage guidance.] Recent \FANDANGO versions support \emph{coverage-guided fuzzing,} optimizing evolutionary input generation towards given execution features~\cite{icst2026-coverage}.
    Integrating \IO grammars into the main line of \FANDANGO will make protocol testing benefit from these extensions.
    \item[Out-of specification testing.] To test ``unhappy paths'',
    \emph{mutating} \IO grammars~\cite{bendrissou2025grammar} will allow generating messages that are \emph{not} covered by the protocol specification.
    \item[Mining protocol specifications.] We are working on tools that extract protocol specifications from existing documentation and implementations;  early experiments with \LLMs yield very promising results~\cite{icst2026-llms}.
    \item[Asynchronous protocols.]
    We want to support for protocols where messages may arrive out of order or concurrently, such as over unreliable or asynchronous channels.
    \item[Time-based behavior.] 
    We want to extend \IO grammars to naturally express time-sensitive behaviors, such as heartbeat messages or timeouts.
    \item[Dynamic participants.] Right now, we assume a fixed set of parties and roles.
    We want to extend them to support dynamic participants, where parties may join or leave during execution.
\end{description}
We look forward to further extending our approach, working with the testing, networking, and modeling research communities towards better protocol testing.

The techniques described herein are implemented as part of the \FANDANGO fuzzer in version~1.1 and later.
\FANDANGO, including a detailed tutorial on protocol fuzzing, is available at
\begin{center}
    \texttt{\url{https://fandango-fuzzer.github.io/}}
\end{center}
The experimental data and the specifications used are available as open source at
\begin{center}
    \texttt{\url{https://github.com/fandango-fuzzer/fandango/tree/io_replication}}
\end{center}

\noindent\textbf{Acknowledgments.}
This work is funded by the European Union (ERC S3, 101093186). Views and opinions expressed are, however, those of the author(s) only and do not necessarily reflect those of the European Union or the European Research Council. Neither the European Union nor the granting authority can be held responsible for them.

\newpage

\bibliographystyle{ACM-Reference-Format}
\bibliography{references}

\appendix

\end{document}